\documentclass[traditabstract]{aa} 
\usepackage{graphicx}
\usepackage{txfonts}
\usepackage{natbib}
\usepackage{color}
\usepackage{lscape}
\usepackage{latexsym}
\bibpunct{(}{)}{;}{a}{}{,} 
\begin{document}
\definecolor{Red}{rgb}{1,0,0}
\authorrunning{Remco F.J. van der Burg et al.}
   \title{Evidence for the inside-out growth of the stellar mass distribution in galaxy clusters since $z\sim 1$}

   \author{Remco~F.J.~van der Burg\inst{1,2}, Henk~Hoekstra\inst{2}, Adam~Muzzin\inst{2,3}, Crist\'obal~Sif\'on\inst{2}, Michael~L.~Balogh\inst{2,4}, Sean~L.~McGee\inst{2,5}}
	   \institute{Laboratoire AIM, IRFU/Service d'Astrophysique - CEA/DSM - CNRS - Universit\'e Paris Diderot, B\^at. 709, CEA-Saclay, 91191 Gif-sur-Yvette Cedex, France\\
                    \email{remco.van-der-burg@cea.fr}        
   \and Leiden Observatory, Leiden University, P.O. Box 9513, 2300 RA Leiden, The Netherlands
\and Kavli Institute for Cosmology, University of Cambridge, Madingley Road, Cambridge, CB3 0HA, United Kingdom
\and Department of Physics and Astronomy, University of Waterloo, Waterloo, Ontario, N2L 3G1, Canada
\and School of Physics and Astronomy, University of Birmingham, Edgbaston, Birmingham, B15 2TT, United Kingdom             }             
             
   \date{Submitted 5 December 2014; Accepted 13 February 2015}

  \abstract {We study the radial number density and stellar mass density distributions of satellite galaxies in a sample of 60 massive clusters at $0.04<z<0.26$ selected from the Multi-Epoch Nearby Cluster Survey (MENeaCS) and the Canadian Cluster Comparison Project (CCCP). In addition to $\sim10,000$ spectroscopically confirmed member galaxies, we use deep $ugri$-band imaging to estimate photometric redshifts and stellar masses, and then statistically subtract fore- and background sources using data from the COSMOS survey. We measure the galaxy number density and stellar mass density distributions in logarithmically spaced bins over 2 orders of magnitude in radial distance from the BCGs. For projected distances in the range $0.1<R/R_{200}<2.0$, we find that the stellar mass distribution is well-described by an NFW profile with a concentration of $c=2.03\pm 0.20$. However, at smaller radii we measure a significant excess in the stellar mass in satellite galaxies of about $10^{11}\,\mathrm{M_{\odot}}$ per cluster, compared to these NFW profiles. We do obtain good fits to generalised NFW profiles with free inner slopes and to Einasto profiles. To examine how clusters assemble their stellar mass component over cosmic time, we compare this local sample to the GCLASS cluster sample at $z\sim 1$, which represents the approximate progenitor sample of the low-$z$ clusters. This allows for a direct comparison, which suggests that the central parts (R$<$0.4 Mpc) of the stellar mass distributions of satellites in local galaxy clusters are already in place at $z\sim 1$, and contain sufficient excess material for further BCG growth. Evolving towards $z=0$, clusters appear to assemble their stellar mass primarily onto the outskirts, making them grow in an inside-out fashion.}  
   \maketitle
%

\hyphenation{in-tra-clus-ter}
\hyphenation{rank-or-der}

\section{Introduction}
Our concordance cosmological model describes a Universe dominated by dark matter and dark energy, in which structures form hierarchically. Within this Lambda-Cold-Dark-Matter ($\Lambda$CDM) framework, N-body simulations provide clear predictions for the structure and evolution of dark matter haloes \citep[e.g.][]{duffy08,dutton14}, and a confrontation with observations provides an important test of our $\Lambda$CDM paradigm. A key open question is how galaxies form in this dark-matter-dominated Universe. The baryonic physics involved may also play a significant role in altering the total mass profiles \citep[e.g.][]{vandaalen11,velliscig14} and therefore complicate a direct comparison with predictions from dark matter simulations. However, as hydrodynamical simulations continue to advance \citep[e.g.][]{schaye10,cen14,genel14,schaye15}, they provide testable predictions of the distribution of baryonic tracers, such as gas and stars. 

An important open question in this context is how well stellar mass traces the underlying dark matter distribution, and if the distribution of galaxies is consistent with what we expect for the sub-haloes in $\Lambda$CDM \citep[e.g.][]{boylankolchin11}. On the scale of our Milky Way, recent hydrodynamical simulations are able to alleviate the tension between the abundance of sub-haloes in N-body simulations, and the observed distribution of satellites, by incorporating baryonic processes such as supernova feedback \citep[e.g.][]{geen13,sawala13}. More massive haloes, such as galaxy clusters, have correspondingly more massive sub-haloes, which is expected to make them more efficient at forming stars, less subjective to feedback processes, and relatively easy to identify through observations. 

Measuring the radial number and stellar mass density distribution of satellite galaxies in clusters has been the focus of several studies. These distributions have been observed to be well described by Navarro-Frenk-White (NFW) \citep{NFW} profiles for group-sized haloes and clusters, from the local Universe to $z\sim 1$ \citep{carlberg97b,lin04,muzzin07,giodini09,budzynski12,vdB14}. Each observational study, however, is based on a different data set and analysis and presents results in a different form. \citet{lin04} and \citet{budzynski12} studied the number density of galaxies, but owing to interactions between galaxies and, in particular, the mass-dependence of the dynamical friction timescale, the number density distribution of galaxies can be different for galaxies with different luminosities or stellar masses. Their results are therefore dependent on the depth of their data set. \citet{giodini09} measured the number density distribution of generally lower mass systems from the COSMOS field. \citet{carlberg97b} and \citet{muzzin07} measured the luminosity density distribution in the $r$-band and K-band, respectively, for clusters from the Canadian Network for Observational Cosmology Survey \citep[CNOC;][]{yee96}. The advantage of this measurement is that, provided the measurements extend significantly below the characteristic luminosity $L^{*}$, it is almost insensitive to the precise luminosity cut. That is because the total luminosity in each radial bin is dominated by galaxies around $L^{*}$. However, especially in the $r$-band, it is not straightforward to relate the luminosity distribution to a stellar mass distribution due to differences in mass-to-light-ratio between different galaxy types, and because the distributions of these types vary spatially. Inconsistencies between all these studies prevent us from drawing firm conclusions on comparisons between them. 

In this paper we present a comprehensive measurement of the radial galaxy number density and stellar mass density from a sample of 60 massive clusters in the local Universe ($0.04<z<0.26$), based on deep $ugri$-band photometry and verified with ample spectroscopic data. The clusters in this sample are approximate descendants of the Gemini Cluster Astrophysics Spectroscopic Survey (GCLASS) cluster sample (by number density), which is a spectroscopic survey of ten rich clusters at $z\sim 1$ \citep[see][]{muzzin12}. In Appendix A, we provide more details on the GCLASS sample selection, and illustrate that the selected clusters make up an approximately representative sample of the high-mass tail of the underlying halo mass distribution. For the GCLASS clusters, the satellite number density (down to galaxies with stellar mass $10^{10.2}\,\mathrm{M_{\odot}}$) and stellar mass density distribution have been measured by \citet[][hereafter vdB14]{vdB14}. By performing the measurements in the present study as consistently as possible with the GCLASS measurement, we study how the stellar mass distribution in massive haloes evolves since $z\sim 1$. 

The structure of this paper is as follows. In Sect.~\ref{sec:dataoverview} we give an overview of the cluster sample, the available spectroscopic data set and the photometric catalogues based on $ugri$-band photometry. Section~\ref{sec:analysis} presents the measurement of the radial density profiles, based on a photometric study but compared with the spectroscopic data for robustness tests. The results are presented in Sect.~\ref{sec:results}, and put into context against low-$z$ literature measurements by a comparison of their best-fitting NFW-profile parameters. In Sect.~\ref{sec:results2} we study the cluster centres in more detail, examining whether the central excess depends on a cluster property in particular. In Sect.~\ref{sec:discussion2} we discuss the observed evolution between $z\sim 1$ and our local study by comparing their stellar mass density profiles on the same physical scale, and discuss the role of build-up of the BCG and ICL components in this context. We summarise and conclude in Sect.~\ref{sec:conclusion}.

All magnitudes we quote are in the AB magnitudes system, and we adopt $\Lambda$CDM cosmology with $\Omega_{\mathrm{m}}=0.3$, $\Omega_{\Lambda}=0.7$ and $\mathrm{H_0=70\, km\, s^{-1}\,  Mpc^{-1}}$. For stellar mass estimates we assume the same Initial Mass Function (IMF) as was used in vdB14, namely the one from \citet{chabrier03}.

\section{Data overview \& processing}\label{sec:dataoverview}
The sample we study consists of 60 massive clusters in the local Universe, drawn from two large X-ray selected surveys: the Multi-Epoch Nearby Cluster Survey (MENeaCS) and the Canadian Cluster Comparison Project (CCCP). For each cluster we acquired deep $ugri$-band photometry (see Sect.~\ref{sec:phot} for details), to allow for a clean cluster galaxy selection.

A substantial number of spectroscopic redshifts in these cluster fields are available from the literature, specifically from CNOC \citep{yee96}, the Sloan Digital Sky Survey Data Release 10 \citep[SDSS DR10;][]{ahn14}, and the Hectospec Cluster Survey \citep[HeCS;][]{rines13}. We searched the NASA/IPAC Extragalactic Database (NED)\footnote{http://ned.ipac.caltech.edu/} to obtain additional spectroscopic information for galaxies that have not been targeted by these surveys, see \citet{sifon14} for details.

\begin{table*}
\caption{The 60 clusters selected from MENeaCS and CCCP that form the basis of this study, with their dynamical properties.}
\label{tab:overview}
\begin{center}
\begin{tabular}{l l r r l l l c c}
\hline
\hline
Name & $z_{\mathrm{spec}}$ &RA$^{\mathrm{a}}$&DEC$^{\mathrm{a}}$& $\sigma_{v}^{\mathrm{b}}$&$M_{200}^{\mathrm{b}}$& $R_{200}^{\mathrm{b}}$ & \multicolumn{2}{c}{Spec-$z$}\\
&&J2000&J2000&$[\mathrm{km/s}]$&[$10^{14}\,\mathrm{M_{\odot}}$]&[Mpc]&Total &Members\\
\hline
       A85&0.055&00:41:50.33&-09:18:11.20&$\,\,\,  967\pm 55$&$10.0\pm 1.7$&$2.0\pm0.1$&  471&  284\\
      A115&0.193&00:55:50.58& 26:24:38.20&$ 1028\pm108$&$11.2\pm 3.5$&$2.0\pm0.2$&  125&   73\\
      A119&0.044&00:56:16.04&-01:15:18.22&$\,\,\,  875\pm 48$&$\,\,\, 7.5\pm 1.2$&$1.9\pm0.1$&  761&  268\\
      A133&0.056&01:02:41.68&-21:52:55.81&$\,\,\,  791\pm 79$&$\,\,\, 5.5\pm 1.7$&$1.7\pm0.2$&   72&   62\\
      A223&0.208&01:37:55.93&-12:49:11.32&$\,\,\,  910\pm 80$&$\,\,\, 7.8\pm 2.1$&$1.8\pm0.2$&   95&   64\\
      A399&0.072&02:57:53.06& 13:01:51.82&$ 1046\pm 47$&$12.5\pm 1.7$&$2.2\pm0.1$&  313&  250\\
      A401&0.074&02:58:57.79& 13:34:57.29&$\,\,\,  933\pm 81$&$\,\,\, 8.9\pm 2.3$&$1.9\pm0.2$&  117&  104\\
      A520&0.201&04:54:14.04& 02:57:09.65&$ 1045\pm 73$&$11.8\pm 2.5$&$2.0\pm0.1$&  278&  153\\
      A521&0.247&04:54:06.86&-10:13:26.01&$ 1002\pm 95$&$10.1\pm 2.9$&$1.9\pm0.2$&  165&   95\\
      A545&0.158&05:32:25.14&-11:32:39.84&$ 1038\pm 89$&$11.8\pm 3.0$&$2.1\pm0.2$&   99&   80\\
      A553&0.067&06:12:41.06& 48:35:44.30&$\,\,\,  665\pm 75$&$\,\,\, 3.3\pm 1.1$&$1.4\pm0.2$&  104&   54\\
      A586&0.170&07:32:20.43& 31:37:57.03&$\,\,\,  803\pm104$&$\,\,\, 5.4\pm 2.1$&$1.6\pm0.2$&  134&   33\\
      A644&0.070&08:17:25.59&-07:30:45.29&$\,\,\,  625\pm 96$&$\,\,\, 2.7\pm 1.2$&$1.3\pm0.2$&   59&   31\\
      A646&0.127&08:22:09.56& 47:05:52.62&$\,\,\,  707\pm 66$&$\,\,\, 3.8\pm 1.1$&$1.4\pm0.1$&  618&  259\\
      A655&0.127&08:25:29.02& 47:08:00.10&$\,\,\,  938\pm 57$&$\,\,\, 8.8\pm 1.6$&$1.9\pm0.1$&  594&  306\\
      A780&0.055&09:18:05.67&-12:05:44.02&$\,\,\,  822\pm113$&$\,\,\, 6.2\pm 2.5$&$1.7\pm0.2$&   42&   33\\
      A795&0.138&09:24:05.30& 14:10:21.00&$\,\,\,  768\pm 59$&$\,\,\, 4.9\pm 1.1$&$1.6\pm0.1$&  330&  166\\
      A961&0.128&10:16:22.93& 33:38:17.98&$\,\,\,  740\pm142$&$\,\,\, 4.4\pm 2.5$&$1.5\pm0.3$&  149&   58\\
      A990&0.142&10:23:39.86& 49:08:38.01&$\,\,\,  829\pm 96$&$\,\,\, 6.1\pm 2.1$&$1.7\pm0.2$&  528&  209\\
     A1033&0.122&10:31:44.31& 35:02:28.71&$\,\,\,  762\pm 52$&$\,\,\, 4.8\pm 1.0$&$1.6\pm0.1$&  496&  170\\
     A1068&0.139&10:40:44.46& 39:57:11.41&$\,\,\,  740\pm160$&$\,\,\, 4.3\pm 2.8$&$1.5\pm0.3$&  621&  104\\
     A1132&0.135&10:58:23.71& 56:47:42.10&$\,\,\,  727\pm 89$&$\,\,\, 4.1\pm 1.5$&$1.5\pm0.2$&  316&  160\\
     A1246&0.192&11:23:58.72& 21:28:48.11&$\,\,\,  956\pm 84$&$\,\,\, 9.1\pm 2.4$&$1.9\pm0.2$&  494&  207\\
     A1285&0.108&11:30:23.79&-14:34:52.79&$\,\,\,  826\pm 90$&$\,\,\, 6.1\pm 2.0$&$1.7\pm0.2$&  168&   77\\
     A1361&0.116&11:43:39.57& 46:21:20.20&$\,\,\,  587\pm 62$&$\,\,\, 2.2\pm 0.7$&$1.2\pm0.1$&  328&  143\\
     A1413&0.142&11:55:18.01& 23:24:17.39&$\,\,\,  881\pm 81$&$\,\,\, 7.3\pm 2.0$&$1.8\pm0.2$&  410&  124\\
     A1650&0.084&12:58:41.52&-01:45:40.90&$\,\,\,  720\pm 48$&$\,\,\, 4.1\pm 0.8$&$1.5\pm0.1$&  787&  266\\
     A1651&0.085&12:59:22.40&-04:11:47.11&$\,\,\,  903\pm 51$&$\,\,\, 8.0\pm 1.4$&$1.9\pm0.1$&  517&  214\\
     A1781&0.062&13:44:52.56& 29:46:15.31&$\,\,\,  419\pm 93$&$\,\,\, 0.8\pm 0.5$&$0.9\pm0.2$&  176&   54\\
     A1795&0.063&13:48:52.58& 26:35:35.81&$\,\,\,  778\pm 51$&$\,\,\, 5.2\pm 1.0$&$1.6\pm0.1$&  508&  191\\
     A1835&0.251&14:01:02.03& 02:52:42.20&$\,\,\,  762\pm106$&$\,\,\, 4.5\pm 1.9$&$1.5\pm0.2$&  690&  195\\
     A1914&0.167&14:25:56.69& 37:48:59.04&$\,\,\,  911\pm 54$&$\,\,\, 7.9\pm 1.4$&$1.8\pm0.1$&  700&  257\\
     A1927&0.095&14:31:06.74& 25:38:00.60&$\,\,\,  725\pm 58$&$\,\,\, 4.2\pm 1.0$&$1.5\pm0.1$&  507&  138\\
     A1942&0.226&14:38:21.86& 03:40:13.22&$\,\,\,  820\pm140$&$\,\,\, 5.6\pm 2.9$&$1.6\pm0.3$&  598&   51\\
     A1991&0.059&14:54:31.50& 18:38:32.71&$\,\,\,  553\pm 45$&$\,\,\, 1.9\pm 0.5$&$1.2\pm0.1$&  613&  175\\
     A2029&0.078&15:10:56.12& 05:44:40.81&$ 1152\pm 58$&$16.6\pm 2.5$&$2.4\pm0.1$&  800&  317\\
     A2033&0.080&15:11:26.55& 06:20:56.40&$\,\,\,  911\pm 69$&$\,\,\, 8.3\pm 1.9$&$1.9\pm0.1$&  608&  190\\
     A2050&0.120&15:16:17.94& 00:05:20.80&$\,\,\,  854\pm 80$&$\,\,\, 6.7\pm 1.9$&$1.7\pm0.2$&  519&  164\\
     A2055&0.103&15:18:45.75& 06:13:55.88&$\,\,\,  697\pm 64$&$\,\,\, 3.7\pm 1.0$&$1.4\pm0.1$&  625&  154\\
     A2064&0.073&15:20:52.23& 48:39:38.81&$\,\,\,  675\pm108$&$\,\,\, 3.4\pm 1.6$&$1.4\pm0.2$&  138&   62\\
     A2065&0.072&15:22:29.16& 27:42:27.00&$ 1095\pm 67$&$14.3\pm 2.6$&$2.3\pm0.1$&  608&  219\\
     A2069&0.114&15:24:08.44& 29:52:54.59&$\,\,\,  966\pm 63$&$\,\,\, 9.7\pm 1.9$&$2.0\pm0.1$&  821&  331\\
     A2104&0.155&15:40:07.85&-03:18:17.03&$ 1081\pm126$&$13.3\pm 4.6$&$2.2\pm0.2$&  194&   90\\
     A2111&0.228&15:39:40.44& 34:25:27.48&$\,\,\,  738\pm 66$&$\,\,\, 4.1\pm 1.1$&$1.4\pm0.1$&  780&  256\\
     A2142&0.090&15:58:20.08& 27:14:01.11&$ 1086\pm 31$&$13.9\pm 1.2$&$2.2\pm0.1$& 1869& 1052\\
     A2163&0.200&16:15:48.97&-06:08:41.64&$ 1279\pm 53$&$21.5\pm 2.7$&$2.5\pm0.1$&  463&  309\\
     A2204&0.151&16:32:46.94& 05:34:33.64&$\,\,\,  782\pm278$&$\,\,\, 5.1\pm 5.4$&$1.6\pm0.6$&  400&  100\\
     A2259&0.160&17:20:09.22& 27:40:10.24&$\,\,\,  901\pm 70$&$\,\,\, 7.7\pm 1.8$&$1.8\pm0.1$&  527&  158\\
     A2261&0.226&17:22:27.16& 32:07:57.36&$\,\,\,  882\pm 86$&$\,\,\, 7.0\pm 2.0$&$1.7\pm0.2$&  604&  206\\
     A2319&0.054&19:21:10.20& 43:56:43.80&$ 1101\pm 99$&$14.7\pm 4.0$&$2.3\pm0.2$&  122&   83\\
     A2409&0.145&22:00:53.51& 20:58:41.80&$\,\,\,  826\pm 94$&$\,\,\, 6.0\pm 2.0$&$1.7\pm0.2$&  341&  101\\
     A2440&0.091&22:23:56.94&-01:34:59.81&$\,\,\,  766\pm 61$&$\,\,\, 4.9\pm 1.2$&$1.6\pm0.1$&  122&   88\\
     A2495&0.079&22:50:19.80& 10:54:13.39&$\,\,\,  631\pm 55$&$\,\,\, 2.8\pm 0.7$&$1.3\pm0.1$&  230&   98\\
     A2597&0.083&23:25:19.70&-12:07:27.70&$\,\,\,  682\pm131$&$\,\,\, 3.5\pm 2.0$&$1.4\pm0.3$&  148&   39\\
     A2670&0.076&23:54:13.60&-10:25:07.50&$\,\,\,  919\pm 46$&$\,\,\, 8.5\pm 1.3$&$1.9\pm0.1$&  400&  241\\
     A2703&0.114&00:05:23.92& 16:13:09.81&$\,\,\,  657\pm 53$&$\,\,\, 3.1\pm 0.8$&$1.4\pm0.1$&  161&   75\\
     MKW3S&0.044&15:21:51.85& 07:42:31.79&$\,\,\,  592\pm 49$&$\,\,\, 2.3\pm 0.6$&$1.2\pm0.1$&  457&  125\\
   RXJ0736&0.118&07:36:38.17& 39:24:51.98&$\,\,\,  432\pm 64$&$\,\,\, 0.9\pm 0.4$&$0.9\pm0.1$&  151&   62\\
  ZWCL0628&0.081&06:31:22.82& 25:01:07.35&$\,\,\,  843\pm 96$&$\,\,\, 6.6\pm 2.2$&$1.8\pm0.2$&  130&   72\\
  ZWCL1023&0.142&10:25:57.99& 12:41:09.31&$\,\,\,  622\pm108$&$\,\,\, 2.6\pm 1.3$&$1.3\pm0.2$&  254&   84\\
  ZWCL1215&0.077&12:17:41.12& 03:39:21.31&$\,\,\,  902\pm 65$&$\,\,\, 8.0\pm 1.7$&$1.9\pm0.1$&  385&  183\\
    \hline
\end{tabular}
\end{center}
\begin{list}{}{}
\item[$^{\mathrm{a}}$] Coordinates of the BCGs. For A2163, a merging system, we take the location of a bright galaxy close to the cluster centre, following the centre adopted by \citet{hoekstra12}. For A115, which consists of a North and a South component, we take the location of the BCG in the Northern part of the system (following \citet{hoekstra12}). 
\item[$^{\mathrm{b}}$] Dynamical properties estimated by \citet{sifon14}.
\end{list}
\end{table*}

\begin{figure}
\resizebox{\hsize}{!}{\includegraphics{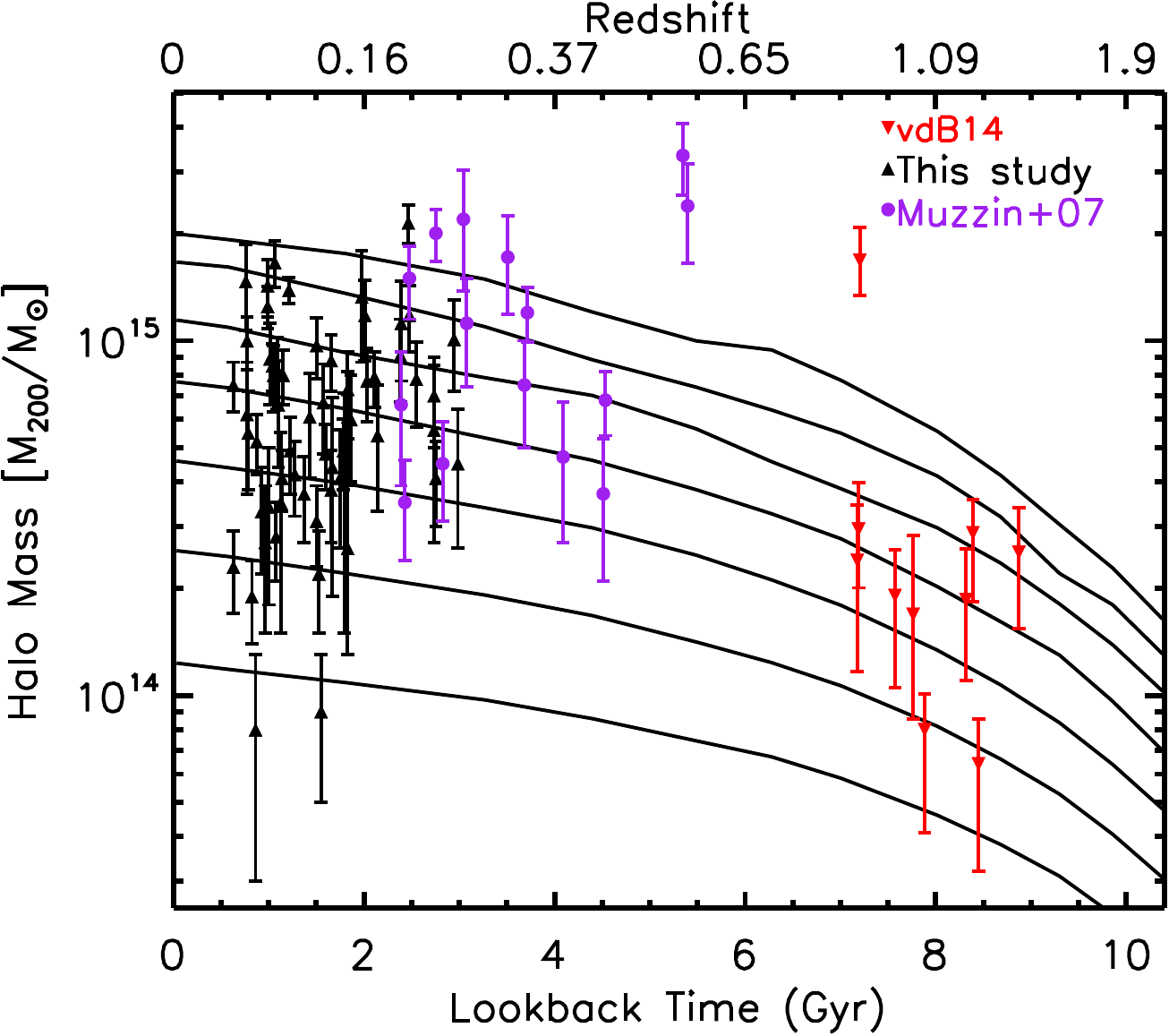}}
\caption{\textit{Lines}: Expected growth curves as a function of cosmic time (or redshift) for massive haloes based on the Millennium simulation \citep{springel05}, in which we followed these haloes at fixed cumulative comoving number density. \textit{Red}: The GCLASS cluster sample studied in vdB14. \textit{Black triangles}: Low redshift cluster sample studied here. \textit{Purple}: The CNOC1 cluster sample studied by \citet{muzzin07}. The cluster samples are linked by the evolutionary growth curves.}
\label{fig:evolution_thesis}
\end{figure}

In addition to the determination of cluster membership these redshifts allow us to estimate dynamical masses \citep{sifon14}. In summary, cluster membership and velocity dispersions are determined using the shifting gapper approach \citep{fadda96}. To relate the velocity dispersion $\sigma_{v}$ to estimates of $R_{200}$, the radius at which the mean interior density is 200 times the critical density ($\rho_{\mathrm{crit}}$), and $M_{200}$, the mass contained within $R_{200}$, the \citet{evrard08} scaling relation is used. We applied the same scaling relation to the GCLASS cluster sample in vdB14 (Sect.~2.1). The 60 clusters used in this study are listed in Table~\ref{tab:overview}.

Figure~\ref{fig:evolution_thesis} suggests that the cluster sample we study covers the mass regime of the likely descendants from GCLASS in the local Universe. Curves in this figure connect haloes selected from the Millennium simulation \citep{springel05} at fixed cumulative comoving number density, and are thus approximate growth curves. The lines are logarithmically spaced, each successive line changing the density by a factor of 3. We also show that the CNOC1 cluster sample, studied by \citet{muzzin07}, are close to the approximate evolutionary sequence, and we will also compare our results to theirs in this paper. In Sect. 5.1 of vdB14, the GCLASS sample is sub-divided to show that SpARCS-1613, which is the highest mass system, has a similar stellar mass distribution as the other clusters, and thus does not bias the analysis in a significant way.

\subsection{Photometry and catalogues in the \textit{ugri}-bands}\label{sec:phot}
Each of these clusters is covered by deep photometric data taken through the $g$-, and $r$-filters using MegaCam mounted on the Canada-France-Hawaii Telescope (CFHT). The data are pre-processed using the \textit{Elixir} pipeline \citep{elixir04}. For MENeaCS, photometric data in the two bands have been taken for these clusters, with a significant dither pattern, and a cadence of several weeks to allow for the detection of type Ia supernovae in these clusters \citep{sand11}. Data for CCCP have been taken consecutively under the best seeing conditions to facilitate weak-lensing measurements \citep{hoekstra12}. For some clusters we further retrieved archival MegaCam data in the $u$-, and $i$-bands (7 and 2 clusters in the respective bands).

The approach we take to process these data further is described in \citet[][hereafter vdB13, Appendix A]{vdB13}, and leads to deep image stacks to measure accurate and precise colours for the purpose of estimating photometric redshifts and stellar masses. We homogenise the PSF of each exposure before stacking, as opposed to homogenising the stack. The former approach leads to a final deep image with a cleaner PSF, especially given that the MENeaCS data have been taken under varying conditions and with substantial dithers. The spatially dependent convolution kernel has been chosen such that the PSF in the final stack has the shape of a circular Gaussian. By applying a Gaussian weight function for aperture fluxes we then optimise colour measurements in terms of S/N \citep[see Appendix A in vdB13;][]{kuijken08}.  

For the clusters that have not been imaged in the $u$-, and $i$-bands with the CFHT, we acquired photometry in these bands using the Wide-Field Camera (WFC), mounted on the Isaac Newton Telescope (INT) in La Palma. Given its field-of-view (FoV) of roughly 30$\times$30 arcmin, we applied a dithered pointing strategy to be able to study the distribution and properties of galaxies that extend up to the clusters' $R_{200}$. Since the angular size of the virial radius depends both on the cluster total mass and its angular diameter distance (through redshift), we varied the number of pointings per cluster. In this way, the area within at least $R_{200}$ is covered to a stacked depth of at least 3 exposures (400s each). Near the cluster centres there are more overlapping pointings which further enhance the depth. 

After pre-processing the images, we convolve them with a position-dependent kernel to homogenise the PSF to a circular Gaussian, similarly to what is done for the MegaCam data. Relative scaling of the photometric zero point between exposures is determined by considering objects that are imaged on overlapping parts between exposures. After these steps, we achieve a systematic uncertainty on flux measurements smaller than 1\% in the two bands. 
 
Because of the excellent image quality and depth in the MegaCam $r$-band stacks, we use these as our detection images. For galaxies with redshift $z\lesssim 0.4$ the $r$-band filter probes the rest-frame SED redward of the $4000\mathring{\mathrm{A}}$ break, which makes the observed $r$-band flux a reasonable proxy for stellar mass. We measure aperture fluxes in the seeing-homogenised images using a Gaussian weight function, which we adjust in size to account for different PSF sizes. To estimate errors on these measurements, we randomly place apertures with the same shape on the seeing-homogenised images and measure the dispersion in the background. Since the flux measurements of our faint sources are background-noise limited, this way we probe the dominant component of the aperture flux error. For the WFC data we compare aperture flux measurements for each source in the individual exposures and (through sigma-clipping) combine this into a flux measurement and error. 

To calibrate the flux measurements in the different filters with respect to each other, we exploit the universal properties of the stellar locus \citep[e.g.][vdB13 (Appendix A)]{slr}. The median limiting magnitudes (5-$\sigma$ for point sources measured with a Gaussian weight function, adjusted in size to accommodate the worst seeing conditions) in the $ugri$-filters are 24.3, 24.8, 24.2 and 23.3, respectively. 

We mask stars brighter than V=15, selected from the Guide Star Catalog II \citep[GSC-II, ][]{lasker08}), and their diffraction spikes and haloes in the images, which typically cover a few percent of the area. The effective area from which we can measure the properties of satellite galaxies is further reduced around and beyond the virial radius, since the fractional area with four-band photometry is reduced. This is especially true for massive clusters at low-$z$, given that they have the largest angular size on the sky. In the following, we take account of these reduced effective areas. 

\section{Analysis}\label{sec:analysis}
Our primary method to measure the radial stellar mass distribution in the ensemble cluster is based on the deep four-band photometry, and relies on a statistical subtraction of background galaxies. We compare this result to the stellar mass distribution of spectroscopically confirmed members as a robustness test. Both approaches are described below. 

\subsection{Statistical background subtraction}\label{sec:analysisbgsub}
The first approach is to estimate a photometric redshift for every galaxy in the cluster images, apply a cut in redshift space ($z<0.3$) and statistically subtract galaxies in the fore-, and background by applying the same redshift cut to the reference COSMOS field. We use $ugri$ photometric data in both our cluster fields and the COSMOS field to estimate photometric redshifts using the EAZY \citep{brammer08} photometric redshift code. We use an $r$-band selected catalogue from the COSMOS field which has been constructed in the same way as the K-band selected catalogue of \citet{muzzin13a}. The field has an effective area of 1.62 $\mathrm{deg^2}$, and we only use data in the $ugri$-filters to provide a fair reference to our cluster sample. 

\begin{figure*}
\resizebox{\hsize}{!}{\includegraphics{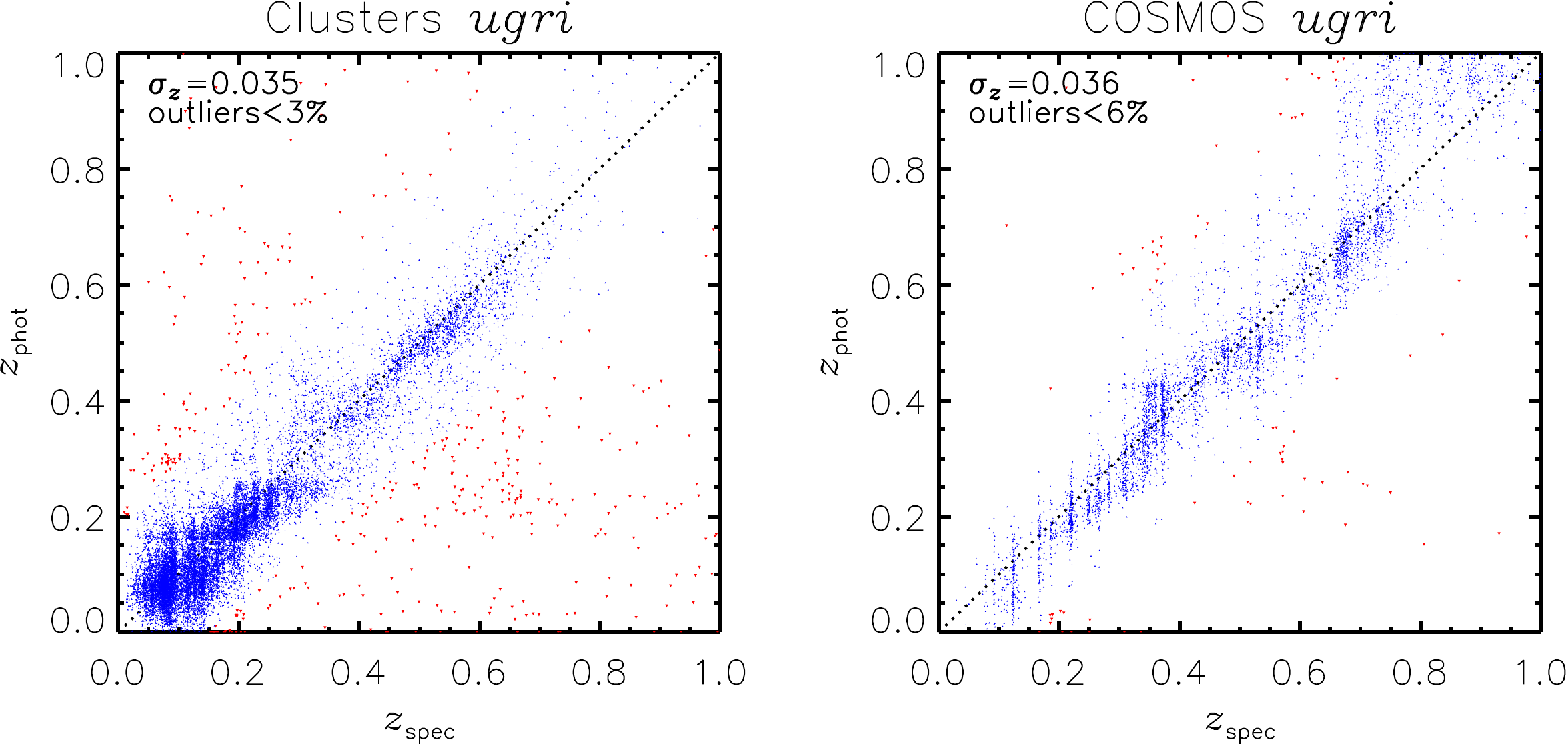}}
\caption{\textit{Left panel}: Spectroscopic versus photometric redshifts for the 60 cluster fields in this study. Outliers, objects for which $\frac{\Delta z}{1+z} > 0.15$, are marked in red. The outlier fraction is less than $3\%$, the scatter (in $\frac{\Delta z}{1+z}$) of the remaining objects is $\sigma_z = 0.035$. \textit{Right panel}: Same for the COSMOS field, also using only the $ugri$-filters. The outlier fraction and scatter are slightly larger as a result of deeper spectroscopic data (in particular at higher redshift where the $ugri$ filters lose their constraining power).}
\label{fig:speczphotz}
\end{figure*}

Because our bluest band is the $u$-band, it is challenging to constrain the location of the $4000 \mathring{\mathrm{A}}$-break for galaxies at low ($z\lesssim 0.15$) redshift, since the break is then located in this filter. Like many redshift codes, EAZY applies a flux-, and redshift-based prior, which gives the redshift probability distribution for a galaxy of a given $r$-band flux P($z$,$r$). This prior has a strong effect in estimating the most probable redshift of a galaxy when the $u$-$g$ colour loses its constraining power (as is the case for redshifts $z\lesssim 0.15$). In the low redshift regime ($z\lesssim0.3$), the comoving volume element $dV_{c}/dz/d\Omega$ is a strong function of redshift \citep[e.g.][]{hogg99}, but the luminosity function does not evolve strongly in this redshift range \citep[e.g.][]{muzzin13b}. Therefore the prior in this regime is decreasing rapidly towards P($z$,$r$)=0 with decreasing redshift, independently of the $r$-band flux. Consequently, according to the prior, it is much more likely to find a galaxy at $z=0.2$ compared to e.g.\ $z=0.1$. Once a field is centred on a massive cluster at low redshift, this prior is no longer applicable since the probability of finding a galaxy to be at the cluster redshift is significantly increased. Besides the general redshift and flux-dependence of the prior, one should therefore also include information on e.g.\ the galaxy's distance to the cluster centre. This however, is beyond our requirements, since we subtract the field statistically, and the volume (and therefore the number of contaminating galaxies) in the field is small for redshift $z<0.3$. A correction on the prior will only affect lower redshifts, and will therefore not change which galaxies survive the redshift cut. For galaxies with a photometric redshift below $z_{\mathrm{EAZY}}=0.16$ we apply a simple correction of the form photo-$z=0.16\cdot(z_{\mathrm{EAZY}}-0.10)/0.06$ to the EAZY output, which we find to lower the scatter between spectroscopic and photometric redshifts for this photometric setup ($ugri$-filters). We apply the same correction to the EAZY output on the COSMOS catalogue. A comparison between spec-$z$'s and photo-$z$'s is shown in in Fig.~\ref{fig:speczphotz}. The axes are truncated at $z$=1.0, but we verified that the confusion between Lyman-break and Balmer-break identifications happens only for $\sim$0.3\% of the total spectroscopic sample, and has a negligible effect on our analysis.

Since the distance modulus is a strong function of redshift in this regime, a small uncertainty in photometric redshift will result in a relatively large uncertainty in luminosity (or stellar mass) of a galaxy. For example, a simple test shows that, for a hypothetical cluster at $z=0.10$, a photo-$z$ bias of +0.005 (-0.005) would result in an inferred luminosity bias that is +11\% (-10\%). For a scatter in the estimated photo-$z$s of $\sigma_{z}=0.035$ (and no bias), we find that the inferred total luminosity in this cluster would be biased high by 19\%. Given that the cluster redshift is well-known, we therefore assign the distance modulus of the cluster to every galaxy in the cluster fields. In order to properly subtract contaminating fore- and background galaxies, we also assign this distance modulus to each galaxy in the reference COSMOS field (after applying the redshift cut). We then use the SED-fitting code FAST \citep{kriek09} to estimate the stellar-mass-to-light ratio (M/L) (in the $r$-band) for each galaxy. For this we again assume the same redshift and distance modulus (corresponding to the cluster) for each galaxy. Then in each of the radial bins (which are scaled by the size $R_{200}$ of each cluster) we measure the area (in angular size) that is covered with four-band photometry, but is not masked by bright stars, and estimate the expected number of sources in this area (which is also different for each cluster through their angular diameter distance) in the COSMOS field. We estimate the total stellar mass and corresponding error for those sources by performing a series of 10,000 Monte-Carlo realisations of the background, by randomly drawing sources from the COSMOS catalogue. We subtract the estimated field values from the raw number counts to obtain the cluster stellar mass density profile. 

It is important to distinguish and account for the different sources of statistical uncertainties that enter our analysis. In the stacked radial profiles, we bootstrap the galaxies in each bin to estimate a statistical error on each data point. We show these error bars in the plots, after including the Poisson uncertainty of the background galaxy counts. We use these errors when fitting profiles, since they are independent between bins, and hence provide a goodness-of-fit test. However, since galaxy clusters are complex systems which are individually not necessarily described by the same profile, we also provide an uncertainty due to sample-to-sample variance. For example, if we would have studied 60 different clusters drawn from the same parent sample (that is, X-ray selected clusters at similar masses and redshifts as the current sample), the resulting stack would have been different. By performing 100 bootstraps (drawing with replacement) of the cluster sample we show that, when stacking a number of 60 clusters with deep photometric data, this sample-to-sample uncertainty dominates over the former statistical error, especially for bins that contain many galaxies and thus have a small statistical error. To estimate this sample-to-sample uncertainty on the best-fitting parameters that describe the stellar mass distribution of the stacked cluster, we perform the fitting procedure on each of the 100 realisations, and combine the range of different best-fitting parameters into an uncertainty. We do not explicitly account for uncertainties on $R_{200}$, but we checked that these have an effect on the data points that is comparable in size to the Poisson uncertainty on the galaxies, and is thus negligible compared to the sample-to-sample uncertainty.

In addition to these statistical uncertainties, and the Poisson noise term in the reference field estimated with the Monte-Carlo realisations, cosmic variance \citep[e.g.][]{somerville04} also contributes to the error in the background. Both the field component that is included in the cluster raw number counts, and the reference field sample from COSMOS, which we subtract from the raw counts, contain this type of uncertainty. However, when several tens of independent cluster fields are stacked, the dominant cosmic variance error arises from the COSMOS reference catalogue. Our analysis, in which we assign the same distance modulus to all galaxies with $z_{\mathrm{phot}} < 0.3$ complicates an estimate of this cosmic variance, since the basic recipes by e.g. \citet{trenti08,moster11} cannot be applied. We do however make an empirical estimate based on catalogues from the 4 spatially independent CFHT Legacy Survey Deep fields \citep{erben09,hildebrandt09a}, which each cover an un-masked area of about 0.8 deg$^2$. After applying the same photometric redshift selection, and masking bright stars, we study the difference between the 4 fields for the following galaxy selections. Assuming a distance modulus corresponding to a redshift of $z=0.15$, the differences in number density of galaxies with stellar mass $10^{9}<\mathrm{M_\star / M_{\odot} < 10^{10}}$ is 14\% among the 4 fields, while the differences for galaxies with stellar mass $\mathrm{M_\star > 10^{10}\, M_{\odot}}$ is about 16\%. When we sum the $r$-band fluxes of all galaxies with $z_{\mathrm{phot}} < 0.3$, as a proxy for the total stellar mass, we find differences between the 4 fields of about 23\% in the total $r$-band flux. Although these fields are a factor of $\sim$2 smaller than the COSMOS field, we will use these differences as a conservative estimate of the cosmic variance error. A measurement of the intrinsic scatter in the profiles of individual clusters requires a more sophisticated investigation of the cosmic variance in annuli centred on individual cluster fields, and is beyond the scope of this paper. 

We perform a consistency check between the COSMOS field and field galaxies that are probed far away from the cluster centres in the low-$z$ cluster data. Although the COSMOS data are significantly deeper, we find no systematic difference in the galaxy stellar mass function between the field probed around the cluster and reference COSMOS field in the regime we are interested in (stellar masses exceeding $\mathrm{M_\star > 10^{9}\, M_{\odot}}$).

To investigate the spatial distribution of individual galaxy types, we locate the red sequence in the ($g-r$)-colour versus $r$-band total magnitude in each of the clusters to distinguish between red and blue galaxies. We find that the slope, and particularly the intersect, of the red sequence vary smoothly with redshift. The dividing line that we use to separate the galaxy types lies just below the red sequence, and is described by $(g-r)_{\mathrm{div}} = [0.475+2.459\cdot z]-[0.036+0.024\cdot z] \times (r_{\mathrm{tot}}-18.0)$, where $z$ is the cluster redshift, and $r_{\mathrm{tot}}$ is the total $r$-band apparent magnitude. As expected, the intersect becomes redder with redshift, wheras the slope becomes steeper. Using the location of spectroscopically confirmed cluster members in colour-magnitude space we finetune the intersect and slope on a cluster-by-cluster basis by hand. This leads to small adjustments with a median absolute difference of 0.017 in the intersect, and a median absolute difference of 0.0016 in the slope, compared to the general equation. In the following we refer to red galaxies as galaxies above the dividing line (which thus lie on the red sequence), and blue galaxies as anything bluer than this dividing line. For each of the clusters we again subtract the field statistically for each of the populations by applying the same colour cut to the COSMOS catalogue. 

\subsection{Comparison with spectroscopic data}
In the method described above, we subtract the galaxies in the fore-, and background statistically based only on the photometric data. However, as discussed in Sect.~\ref{sec:dataoverview}, we can use a substantial number of spectroscopic redshifts in the cluster fields from the literature. In this second approach we measure the stellar mass contained in spectroscopically confirmed cluster members to provide a lower limit to the full stellar mass distribution.

\begin{figure}
\resizebox{\hsize}{!}{\includegraphics{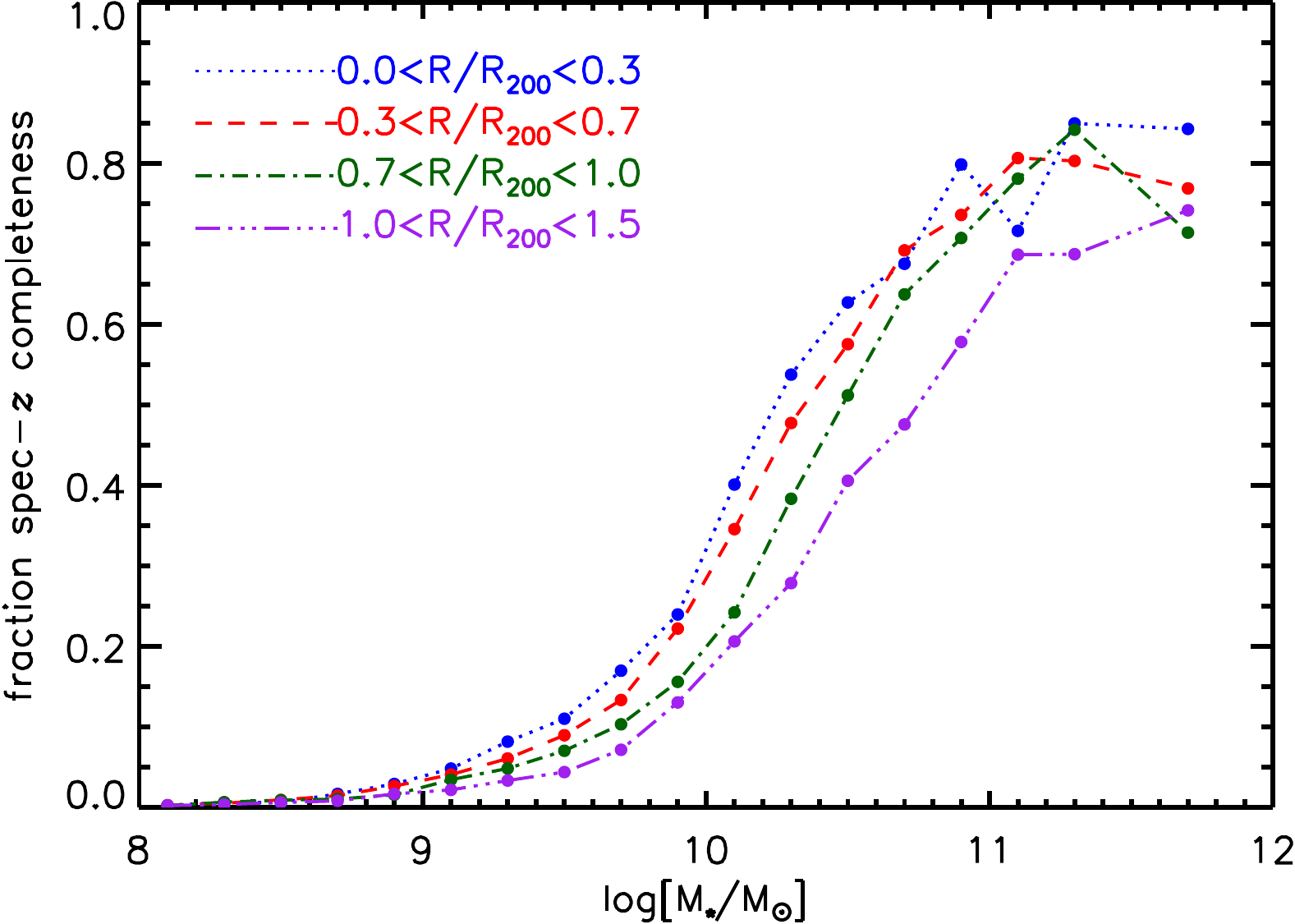}}
\caption{Spectroscopic completeness for sources with a photometric redshift $z < 0.3$ as a function of stellar mass (assuming the same distance modulus as the cluster redshift). The four lines show different radial bins. For targets of a given stellar mass, the spectroscopic completeness is slightly higher for those that are closer to the cluster centres. For each of the radial bins, the completeness is larger than 70\% for stellar masses $\mathrm{M_\star > 10^{11}\, M_{\odot}}$.}
\label{fig:specz_compl}
\end{figure}

Since the spectroscopic data set is obtained after combining several different surveys, the way the spectroscopic targets have been selected is not easily reconstructed. Fig.~\ref{fig:specz_compl} shows the spectroscopic completeness for all galaxies with a photometric redshift $z<0.3$ as a function of stellar mass (assuming the same distance modulus as the cluster redshift), and for different radial bins. For stellar masses $\mathrm{M_\star > 10^{11}\, M_{\odot}}$, the completeness is high ($>70\%$) in each of the radial bins. Since these objects constitute most of the total stellar mass distribution (see vdB14 (Fig.~2) for this argument), we can get a fairly complete census of stellar mass by just considering the galaxies for which we have a spectroscopic redshift. We estimate the fraction of the stellar mass that is in spectroscopically confirmed members, for each of the four radial bins. For this we assume a stellar mass distribution following a \citet{schechter76} function with characteristic mass $M^{*}=10^{11}\,\rm{M_{\odot}}$, and low-mass slope $\alpha = -1.3$. These choices are motivated by the low-$z$ bin of the field stellar mass function as measured by \citet{muzzin13b}. When we multiply this distribution with the completeness curves as shown in Fig.~\ref{fig:specz_compl}, we find a spectroscopic completeness for the total stellar mass in satellite galaxies of 59\%, 57\%, 52\%, and 43\% for the four radial bins, respectively. 

\subsection{The presence of the BCG}
In order to measure the number density and stellar mass density profiles of the satellites close to the cluster centres, we subtract the primary component of the BCGs' flux-profiles with {\tt GALFIT} \citep{galfit} prior to source extraction and satellite photometry. We find that this step has a significant impact on the measured number density of faint satellites ($\mathrm{M_\star < 10^{10}\, M_{\odot}}$) near the cluster centres, which was also mentioned by \citet{budzynski12}. In the next section we mask the inner two bins (for which $R < 0.02 \cdot R_{200}$) given that their values change by more than two times their statistical error. We find that the effect on the number density distribution of more massive satellites ($\mathrm{M_\star > 10^{10}\, M_{\odot}}$) is negligible. The effect is largest in the first logarithmic bin (for which $R \approx 0.015 \cdot R_{200}$), but even here the results change by less than the size of the statistical error. The effects on the stellar mass density distribution are also smaller than the statistical error. The reason for this is that the stellar mass density distribution is primarily composed of more massive satellites which are relatively unobscured by the BCG. We therefore conclude that, although we remove the BCG profile prior to satellite detection and photometry, doing so has a negligible effect on the measured stellar mass density profile. 

\section{Results in the context of the NFW profile}\label{sec:results}
In this section we present the galaxy number and stellar mass density distributions of the 60 clusters we study, based on the two independent analyses described in Sect.~\ref{sec:analysis}. We discuss these results by considering the NFW \citep{NFW} fitting function, since that is the parameterisation generally used in previous studies. We can therefore compare the results in this context with measurements in the literature, both at low and high redshift. 

\subsection{Galaxy number density profile}
\begin{figure*}[t]
\resizebox{\hsize}{!}{\includegraphics{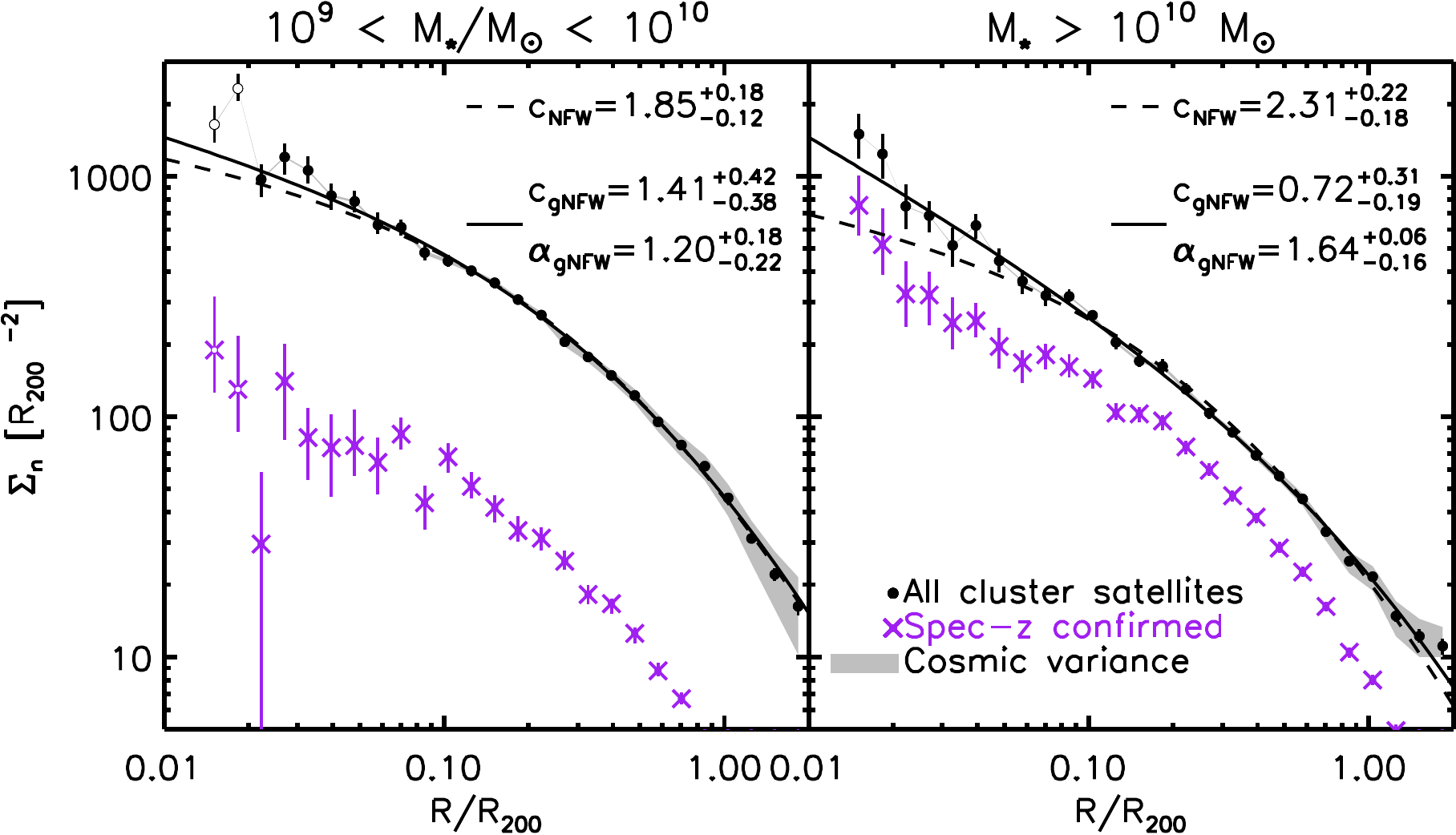}}
\caption{Galaxy number density distributions for masses $10^{9}<\mathrm{M_\star / M_{\odot} < 10^{10}}$ (left panel), and $\mathrm{M_\star > 10^{10}\, M_{\odot}}$ (right panel) for the ensemble cluster at $z\sim 0.15$. Black points with the best-fitting projected NFW (dashed) and gNFW (solid) functions are our best estimates for the cluster number counts. The inner two points in the left panel are masked due to obscuration from the BCG, which is more severe for low-mass galaxies, and are excluded from the fitting. Purple points indicate the number of spectroscopically confirmed cluster members.}
\label{fig:number_profile_ensemble}
\end{figure*}
Ignoring baryonic physics, the galaxy number density distribution in cluster haloes can be compared to the distribution of sub-haloes in N-body simulations as a test of $\Lambda$CDM. Due to mergers and interactions between galaxies, and in particular the mass-dependence of the dynamical friction timescale, the number density distribution of galaxies may be different for galaxies with different stellar masses. 

Figure~\ref{fig:number_profile_ensemble} shows the projected galaxy number density distribution for galaxies with stellar masses $10^{9}<\mathrm{M_\star / M_{\odot} < 10^{10}}$ (left panel), and $\mathrm{M_\star > 10^{10}\, M_{\odot}}$ (right panel) in the ensemble cluster. Before stacking the 60 clusters, their radial distances to the BCGs are scaled by $R_{200}$, but the BCGs themselves are not included in the data points. Error bars reflect bootstrapped errors arising from both the cluster galaxy counts and the field value that is subtracted. The shaded area around the data points shows the systematic effect due to cosmic variance in the background, which we estimated in Sect.~\ref{sec:analysisbgsub}. The number of spectroscopically confirmed cluster members follow a similar distribution but have a different normalisation due to spectroscopic incompleteness. 

We fit projected NFW profiles to the data points, and show those corresponding to the minimum $\chi^2$ values with the dashed lines in Fig.~\ref{fig:number_profile_ensemble}. For the lower-mass galaxies ($10^{9}<\mathrm{M_\star / M_{\odot} < 10^{10}}$), we find an overall goodness-of-fit of $\chi^2/d.o.f.=1.19$, with a concentration of $c=1.85^{+0.18+0.09}_{-0.12-0.09}$. Both a sample-to-sample variance (first) and systematic (second) error are quoted. For the higher-mass galaxies ($\mathrm{M_\star > 10^{10}\, M_{\odot}}$), the overall goodness-of-fit is $\chi^2/d.o.f.=3.00$ with a concentration of $c=2.31^{+0.22+0.32}_{-0.18-0.29}$. In both stellar mass bins, we find that the best-fitting NFW function gives a reasonable description of the data for most of the cluster ($R \gtrsim 0.10 \cdot R_{200}$), but that the centre has an excess in the number of galaxies compared to the NFW profile. In the next section we provide a more detailed investigation of this excess; in this section we continue working with the standard NFW profile in order to compare with previous work.

The number density and luminosity density profiles of group and cluster sized haloes in the literature have generally been measured on smaller samples, and do not focus on the smallest radial scales around the BCGs. On the scales these studies have focussed on, NFW profiles have been shown to be an adequate fit to the data over the whole radial range. We therefore compare the concentration parameters fitted by the NFW profile with the values presented in the literature. 

\citet{lin04} studied the average number density profile of a sample of 93 clusters at $0.01<z<0.09$ with 2MASS K-band data. They were able to measure down to a magnitude limit (Vega) of $\mathrm{K_{s, lim}}=13.5$, which corresponds to $\mathrm{M_\star \approx 10^{10}\, M_{\odot}}$ at $z=0.05$ \citep{bell01}. Although they studied clusters with a lower mass range than we probe, they found a number density concentration of $c=2.90^{+0.21}_{-0.22}$, which is comparable to the value that we find for the high mass galaxies ($\mathrm{M_\star > 10^{10}\, M_{\odot}}$).

\citet{budzynski12} measured the radial distribution of satellite galaxies in groups and clusters in the range $0.15<z<0.40$ from the SDSS DR7. For the satellite galaxies they applied a magnitude limit of $M_{r}=-20.5$. This corresponds to about $\mathrm{M_\star = 10^{10.5}\, M_{\odot}}$ for galaxies with a high M/L. The best-fitting concentration parameter of $c\sim2.6$ they found is also consistent with our measurement for the high-mass sample. They found that the concentration of the satellite distribution decreases slightly as their brightness increases, but note that they compared satellites in a higher luminosity range with respect to our study.

vdB14 measured the number density distribution of the GCLASS cluster ensemble at $z\sim 1$ down to a stellar mass of $\mathrm{M_\star = 10^{10.2}\, M_{\odot}}$. They measured an NFW concentration parameter of $c=5.14^{+0.54}_{-0.63}$, which is significantly higher than the value we find for the low-$z$ sample, indicating that there is a substantial evolution with redshift. A comparison between the number density distribution and the stellar mass density distribution presented in vdB14 suggests that the more massive galaxies are situated closer towards the cluster centres than lower mass galaxies, which is qualitatively consistent with the trend we find here.

\subsection{Stellar mass density profile}
\begin{figure}
\resizebox{\hsize}{!}{\includegraphics{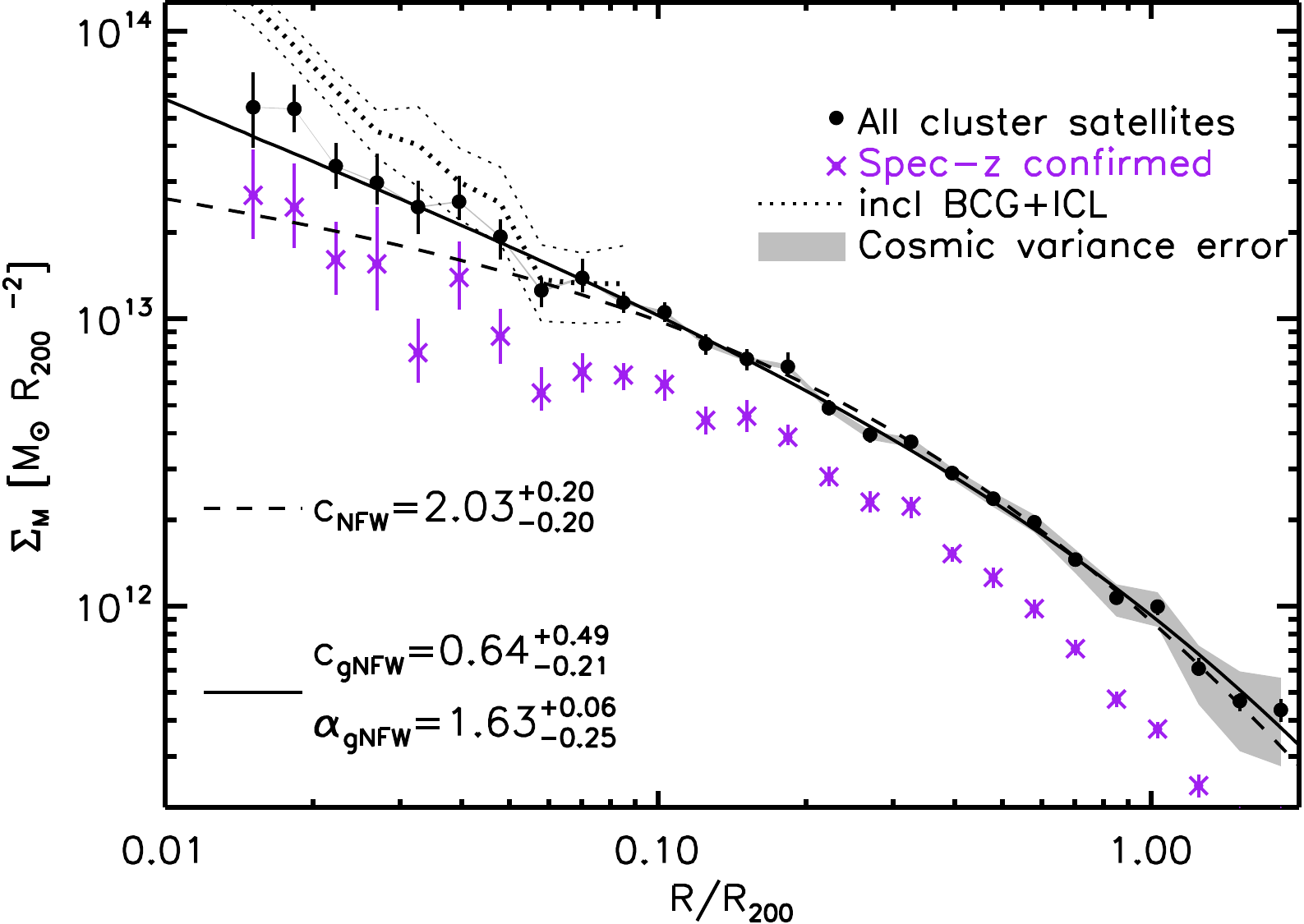}}
\caption{Stellar mass density distribution of the ensemble cluster at $z\sim 0.15$. \textit{Black points:} cluster stellar mass distribution, with best-fitting projected NFW (dashed) and gNFW (solid) functions. \textit{Purple points:} stellar mass distribution in spectroscopically confirmed cluster members. \textit{Dotted line:} total stellar mass density profile on the images (background subtracted, with the 68\% confidence region around these values shown by the thinner dotted lines), including the BCG and part of the ICL component.}
\label{fig:profile_ensemble}
\end{figure}
Whereas the number density distribution of galaxies depends sensitively on the stellar mass range considered (or the depth of the data set), the stellar mass density distribution is less sensitive to this, because it is primarily set by the distribution of galaxies around the characteristic mass ($M^*$). Figure~\ref{fig:profile_ensemble} shows the radial stellar mass density distribution of satellite galaxies in the ensemble cluster. Radial distances are normalised by the clusters' $R_{200}$. Black data points give the background-subtracted cluster stellar mass distribution, with errors estimated by bootstrapping the galaxies in the stack. Ignoring systematic uncertainties such as the shape of the IMF, stellar mass errors of individual galaxies are negligible compared to this bootstrap error. We do, however, show a systematic uncertainty of 23\% in the background due to cosmic variance by the shaded region around the data points. The spectroscopic completeness in terms of total stellar mass is about 50\%, and does not significantly depend on radial distance (cf.\ Fig.~\ref{fig:specz_compl}).

As for the number density profiles, we fit a projected NFW profile to the black data points, minimizing the $\chi^2$ value. Again we find that the best-fitting NFW function gives a reasonable description (dashed curve in the figure) of the data for most of the cluster ($R \gtrsim 0.10 \cdot R_{200}$), but that the central parts show a significant excess of stellar mass in satellites near the centre compared to this function. To provide a consistent comparison with previous studies, we consider the best-fitting NFW concentration parameter, $c=2.03^{+0.20+0.60}_{-0.20-0.40}$, here and present a more detailed investigation of the central excess in the next section. As before, sample-to-sample variance (first) and systematic errors (from cosmic variance in the background, second) are quoted. We again limit a comparison with literature studies to the NFW concentration parameters, since the NFW profile has been used as an adequate description of previous measurements.

\citet{muzzin07} measure the K-band luminosity profiles for a stack of 15 CNOC1 \citep{yee96} clusters in the redshift range $0.2<z<0.5$. In this redshift range, the luminosity in the K-band is expected to be a good proxy for stellar mass. They find a concentration of the luminosity density of $c=4.28\pm 0.70$. These clusters are only slightly more massive than the progenitors of the sample we study (see Fig.~\ref{fig:evolution_thesis}). 

At higher redshift ($z\sim 1$), vdB14 present the stellar mass density distribution of the GCLASS cluster sample, and find that an NFW profile with a high concentration of $c=7.12^{+1.53}_{-0.99}$ fits the data. These systems are likely to grow into the low-$z$ clusters studied in this paper. 

Together, these studies span an interval of about 8 Gyr of cosmic time, and comparisons among these results indicate that the stellar mass distribution in clusters evolves significantly. This trend is visualised by the four data points in Fig.~\ref{fig:c_evolution_new}, which represent the studies discussed above. The black points represent the present study, divided over two redshift bins (see Sect.~\ref{sec:samplesplits}).

\subsubsection{Discussion}\label{sec:discussion1}
Although satellite galaxies are expected to mark the location of dark matter sub-haloes, a comparison with theoretical predictions has limitations. Most studies are based on large N-body simulations \citep{springel05}, and dark matter haloes falling into larger haloes experience tidal forces leading to the stripping of their constituent particles \citep{ghigna00,binney08}, also see \citet{natarajan02,gillis13} for observational studies. As a sub-halo falls into the main halo, it will continuously lose mass through the process of tidal stripping, and it may eventually fall below the mass resolution of the simulation. The sub-halo is then no longer identified as such, its mass is deposited on the central galaxy or dispersed between the galaxies, and its orbit is no longer defined. For this reason, the radial distribution of sub-haloes is less concentrated than the dark matter in N-body simulations \citep{nagai05}. While the sub-haloes in these dissipationless simulations are eventually destroyed, the galaxies that have formed inside of them are expected to be more resistive to tidal forces. The observed distribution of satellite galaxies may therefore be more similar to the distribution of dark matter in N-body simulations. 

\begin{figure}
\resizebox{\hsize}{!}{\includegraphics{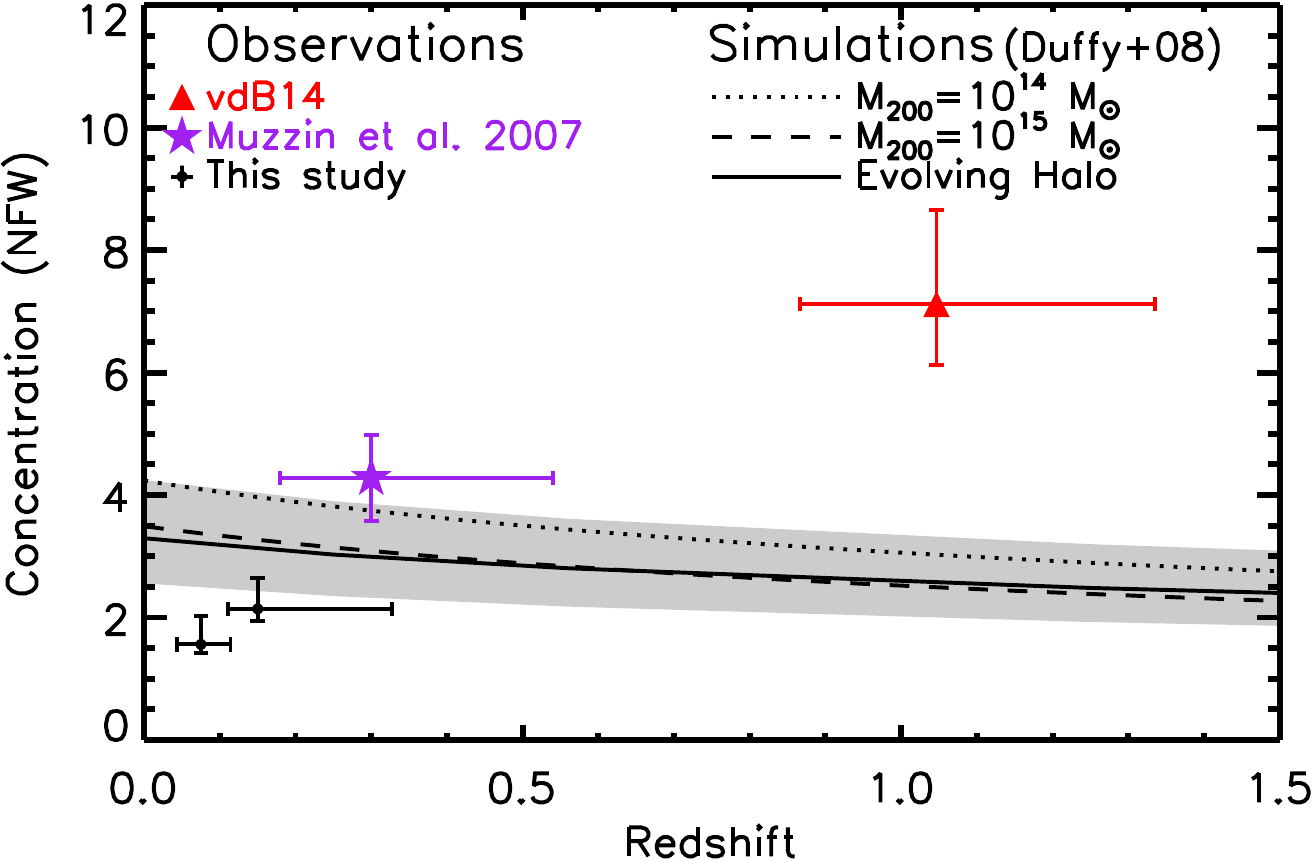}}
\caption{\textit{Black points}: Stellar mass density concentration for the clusters used in this study, split in two redshift bins. \textit{Purple}: K-band luminosity density concentration in CNOC1 from \citet{muzzin07}. \textit{Red}: Stellar mass density concentration in GCLASS from vdB14. The horizontal bars indicate the redshift range for each sample. \textit{Black lines}: The NFW concentration in the sample of relaxed haloes from \citet{duffy08} as a function of redshift. \textit{Dotted and dashed}: Haloes of a given mass as a function of redshift. \textit{Solid}: NFW concentration of a halo that is evolving in mass, with scatter given by the shaded region.}
\label{fig:c_evolution_new}
\end{figure}

In Fig.~\ref{fig:c_evolution_new}, the evolution of the NFW concentration parameter in N-body simulations is shown \citep{duffy08}, for relaxed haloes with masses similar to the massive galaxy clusters studied here. The grey area indicates the intrinsic scatter around the solid line. For the full sample of haloes (including non-relaxed), the average concentration is slightly lower, but the intrinsic scatter is slightly larger. But note that these results are based on a WMAP5 cosmology, and e.g. \citet{maccio08} have shown that the shapes of dark matter halo profiles depend sensitively on the cosmological parameters. Indeed, \citet{dutton14} base their study on a \textit{Planck} cosmology, which is characterised by a larger $\Omega_{\mathrm{m}}$ and $\sigma_8$, and find that the haloes are more concentrated by about 30\% in our mass and redshift regime. The concentration shown in Fig.~\ref{fig:c_evolution_new} might therefore be 30\% higher, but the evolution in the concentration parameters is qualitatively independent of the cosmology. 

Therefore, the evolution of the simulated dark matter distribution is significantly different than the observed evolution of the stellar mass distribution. Note again that, at low redshift we find that the NFW profile does not give a proper representation of the data at the inner parts of the cluster. In the next section we focus on the excess of stellar mass in more detail, and expand the discussion in Sect.~\ref{sec:discussion2}.

\begin{table*}
\caption{Parameters describing the best-fitting NFW (where $\alpha$ is fixed to 1) and gNFW (where $\alpha$ is a free parameter) profiles to the radial density distributions. Both random (due to sample-to-sample variance) and systematic (due to cosmic variance) uncertainties are quoted. The reduced $\chi^2$ values show that the gNFW profiles give significantly better descriptions of the data. The sample is split according to several criteria to investigate the dependence of these parameters on cluster properties.}
\label{tab:nfwpars}
\begin{center}
\begin{tabular}{l | c c c | c c c | c c c}
\hline
\hline
&\multicolumn{3}{c}{Stellar Mass}&\multicolumn{3}{c}{Number density}&\multicolumn{3}{c}{Number density}\\
&\multicolumn{3}{c}{density}&\multicolumn{3}{c}{$10^{9}<\mathrm{M_\star / M_{\odot} < 10^{10}}$}&\multicolumn{3}{c}{$\mathrm{M_\star > 10^{10}\, M_{\odot}}$}\\
Sample&$c_{\mathrm{gNFW}}$&$\alpha_{\mathrm{gNFW}}$&$\frac{\chi^2}{d.o.f.}$&$c_{\mathrm{gNFW}}$&$\alpha_{\mathrm{gNFW}}$&$\frac{\chi^2}{d.o.f.}$&$c_{\mathrm{gNFW}}$&$\alpha_{\mathrm{gNFW}}$&$\frac{\chi^2}{d.o.f.}$\\
\hline
All (NFW) &$2.03^{+0.20+0.60}_{-0.20-0.40}$&1 (fixed)&2.51&$1.85^{+0.18+0.09}_{-0.12-0.09}$&1 (fixed)&1.19&$2.31^{+0.22+0.32}_{-0.18-0.29}$&1 (fixed)&3.00\\
All&$0.64^{+0.49+0.73}_{-0.21-0.33}$&$1.63^{+0.06+0.09}_{-0.25-0.17}$&1.24&$1.41^{+0.42+0.18}_{-0.38-0.12}$&$1.20^{+0.18+0.03}_{-0.22-0.04}$&1.04&$0.72^{+0.31+0.31}_{-0.19-0.28}$&$1.64^{+0.06+0.08}_{-0.16-0.06}$&1.19\\
$z<0.114$&$0.36^{+0.76+0.14}_{-0.09-0.06}$&$1.66^{+0.06+0.00}_{-0.35-0.01}$&1.06&$0.70^{+0.52+0.06}_{-0.28-0.04}$&$1.38^{+0.13+0.01}_{-0.17-0.01}$&1.09&$0.71^{+0.61+0.16}_{-0.29-0.09}$&$1.63^{+0.09+0.01}_{-0.22-0.02}$&0.83\\
$z\geq0.114$ &$0.97^{+1.57+1.67}_{-0.33-0.62}$&$1.50^{+0.11+0.20}_{-0.52-0.41}$&1.71&$1.47^{+1.07+0.38}_{-0.53-0.21}$&$1.28^{+0.20+0.05}_{-0.40-0.11}$&1.06&$1.09^{+0.85+0.69}_{-0.55-0.46}$&$1.50^{+0.18+0.14}_{-0.32-0.17}$&1.97\\
$M_{200} < 8.6 \cdot 10^{14}\,\mathrm{M_{\odot}}$&$0.31^{+0.72+0.31}_{-0.18-0.20}$&$1.68^{+0.09+0.07}_{-0.25-0.08}$&1.06&$0.64^{+0.49+0.06}_{-0.21-0.10}$&$1.45^{+0.18+0.03}_{-0.22-0.01}$&0.64&$0.19^{+0.74+0.26}_{-0.09-0.10}$&$1.84^{+0.05+0.02}_{-0.31-0.08}$&1.63\\
$M_{200}\geq 8.6 \cdot 10^{14}\,\mathrm{M_{\odot}}$&$1.30^{+0.67+1.20}_{-0.43-0.57}$&$1.42^{+0.10+0.14}_{-0.22-0.24}$&1.67&$2.10^{+1.37+0.29}_{-0.73-0.21}$&$1.00^{+0.30+0.05}_{-0.50-0.08}$&2.08&$1.14^{+0.43+0.47}_{-0.47-0.31}$&$1.48^{+0.12+0.07}_{-0.18-0.10}$&0.84\\
$M_\mathrm{{\star,BCG}} < 9.1 \cdot 10^{11}\,\mathrm{M_{\odot}}$&$1.09^{+1.48+1.49}_{-0.42-0.64}$&$1.40^{+0.14+0.20}_{-0.61-0.34}$&2.17&$2.30^{+1.17+0.40}_{-0.63-0.28}$&$1.01^{+0.18+0.06}_{-0.32-0.10}$&1.31&$0.86^{+0.71+0.60}_{-0.49-0.40}$&$1.57^{+0.12+0.12}_{-0.28-0.15}$&2.12\\
$M_\mathrm{{\star,BCG}}\geq 9.1 \cdot 10^{11}\,\mathrm{M_{\odot}}$&$0.53^{+0.48+0.46}_{-0.13-0.22}$&$1.61^{+0.07+0.07}_{-0.16-0.12}$&0.82&$0.83^{+0.48+0.06}_{-0.22-0.07}$&$1.34^{+0.21+0.02}_{-0.09-0.01}$&1.04&$0.66^{+0.55+0.21}_{-0.15-0.19}$&$1.64^{+0.11+0.05}_{-0.19-0.04}$&0.56\\
$n_{1\mathrm{Mpc}, M_{\star}>10^{10}\,\mathrm{M_{\odot}}} < 87$&$0.89^{+0.60+0.70}_{-0.40-0.33}$&$1.30^{+0.15+0.09}_{-0.35-0.17}$&0.87&$0.91^{+0.48+0.12}_{-0.42-0.12}$&$1.35^{+0.10+0.04}_{-0.30-0.03}$&0.70&$0.54^{+0.45+0.33}_{-0.35-0.18}$&$1.52^{+0.13+0.06}_{-0.27-0.10}$&0.80\\
$n_{1\mathrm{Mpc}, M_{\star}>10^{10}\,\mathrm{M_{\odot}}} \geq 87$&$0.92^{+0.70+0.97}_{-0.40-0.47}$&$1.56^{+0.12+0.13}_{-0.18-0.22}$&1.51&$1.96^{+0.96+0.26}_{-0.84-0.21}$&$1.05^{+0.23+0.06}_{-0.37-0.06}$&1.75&$1.02^{+0.90+0.60}_{-0.20-0.30}$&$1.62^{+0.06+0.06}_{-0.24-0.14}$&1.18\\
$K_0 < \mathrm{70\,keV\,cm^2}$&$1.64^{+1.21+0.76}_{-0.59-0.43}$&$1.01^{+0.23+0.11}_{-0.57-0.19}$&1.46&$1.70^{+0.75+0.18}_{-0.55-0.12}$&$0.88^{+0.19+0.03}_{-0.29-0.05}$&1.90&$0.89^{+0.86+0.34}_{-0.27-0.21}$&$1.50^{+0.14+0.05}_{-0.36-0.09}$&0.71\\
$K_0 \geq \mathrm{70\,keV\,cm^2}$&$0.57^{+1.06+0.76}_{-0.34-0.37}$&$1.72^{+0.11+0.10}_{-0.29-0.17}$&1.06&$2.19^{+1.34+0.33}_{-1.06-0.20}$&$0.98^{+0.35+0.05}_{-0.25-0.08}$&2.00&$0.80^{+1.23+0.52}_{-0.37-0.28}$&$1.67^{+0.16+0.06}_{-0.24-0.12}$&0.97\\
\hline
\end{tabular}
\end{center}
\end{table*}

\begin{table*}
\caption{Similar to Table~\ref{tab:nfwpars}, but showing the parameters corresponding to the best-fitting Einasto profiles.}
\label{tab:einastopars}
\begin{center}
\begin{tabular}{l | c c c | c c c | c c c}
\hline
\hline
&\multicolumn{3}{c}{Stellar Mass}&\multicolumn{3}{c}{Number density}&\multicolumn{3}{c}{Number density}\\
&\multicolumn{3}{c}{density}&\multicolumn{3}{c}{$10^{9}<\mathrm{M_\star / M_{\odot} < 10^{10}}$}&\multicolumn{3}{c}{$\mathrm{M_\star > 10^{10}\, M_{\odot}}$}\\
Sample&$c_{\mathrm{EIN}}$&$\alpha_{\mathrm{EIN}}$&$\frac{\chi^2}{d.o.f.}$&$c_{\mathrm{EIN}}$&$\alpha_{\mathrm{EIN}}$&$\frac{\chi^2}{d.o.f.}$&$c_{\mathrm{EIN}}$&$\alpha_{\mathrm{EIN}}$&$\frac{\chi^2}{d.o.f.}$\\
\hline
All&$1.96^{+0.27+0.70}_{-0.43-0.75}$&$0.11^{+0.07+0.04}_{-0.03-0.06}$&1.40&$1.73^{+0.20+0.11}_{-0.10-0.06}$&$0.21^{+0.07+0.01}_{-0.03-0.01}$&1.03&$2.31^{+0.32+0.39}_{-0.28-0.43}$&$0.10^{+0.02+0.02}_{-0.02-0.03}$&1.16\\
$z<0.114$&$1.03^{+0.79+0.54}_{-0.35-0.36}$&$0.07^{+0.14+0.01}_{-0.02-0.02}$&1.23&$1.14^{+0.38+0.09}_{-0.52-0.08}$&$0.16^{+0.05+0.01}_{-0.05-0.00}$&1.17&$2.20^{+0.32+0.31}_{-0.58-0.26}$&$0.11^{+0.10+0.01}_{-0.02-0.01}$&0.97\\
$z\geq0.114$ &$2.07^{+0.37+0.61}_{-0.33-0.78}$&$0.14^{+0.04+0.06}_{-0.06-0.08}$&1.73&$2.04^{+0.30+0.11}_{-0.30-0.12}$&$0.18^{+0.02+0.01}_{-0.10-0.02}$&1.09&$2.37^{+0.47+0.35}_{-0.43-0.40}$&$0.13^{+0.05+0.03}_{-0.05-0.03}$&1.89\\
$M_{200} < 8.6 \cdot 10^{14}\,\mathrm{M_{\odot}}$&$0.89^{+0.64+0.77}_{-0.37-0.65}$&$0.07^{+0.16+0.03}_{-0.02-0.03}$&1.00&$1.20^{+0.33+0.10}_{-0.27-0.12}$&$0.14^{+0.09+0.01}_{-0.02-0.01}$&0.69&$1.52^{+0.81+0.83}_{-0.44-0.65}$&$0.05^{+0.08+0.01}_{-0.01-0.03}$&1.47\\
$M_{200}\geq 8.6 \cdot 10^{14}\,\mathrm{M_{\odot}}$&$2.36^{+0.21+0.38}_{-0.49-0.53}$&$0.17^{+0.03+0.04}_{-0.07-0.08}$&1.77&$1.94^{+0.13+0.08}_{-0.27-0.07}$&$0.26^{+0.04+0.02}_{-0.06-0.01}$&1.74&$2.32^{+0.35+0.31}_{-0.45-0.30}$&$0.15^{+0.05+0.02}_{-0.05-0.03}$&0.94\\
$M_\mathrm{{\star,BCG}} < 9.1 \cdot 10^{11}\,\mathrm{M_{\odot}}$&$1.90^{+0.37+0.60}_{-0.43-0.71}$&$0.16^{+0.13+0.06}_{-0.07-0.09}$&2.23&$2.16^{+0.21+0.09}_{-0.29-0.12}$&$0.24^{+0.05+0.01}_{-0.05-0.02}$&1.22&$2.20^{+0.37+0.45}_{-0.53-0.46}$&$0.12^{+0.07+0.03}_{-0.03-0.03}$&2.11\\
$M_\mathrm{{\star,BCG}}\geq 9.1 \cdot 10^{11}\,\mathrm{M_{\odot}}$&$1.49^{+0.42+0.59}_{-0.38-0.47}$&$0.10^{+0.15+0.02}_{-0.01-0.04}$&0.85&$1.27^{+0.24+0.09}_{-0.26-0.05}$&$0.17^{+0.08+0.00}_{-0.02-0.01}$&1.13&$2.11^{+0.40+0.36}_{-0.40-0.35}$&$0.10^{+0.15+0.01}_{-0.01-0.02}$&0.52\\
$n_{1\mathrm{Mpc}, M_{\star}>10^{10}\,\mathrm{M_{\odot}}} < 87$&$1.32^{+0.11+0.46}_{-0.43-0.45}$&$0.19^{+0.06+0.05}_{-0.14-0.08}$&0.98&$1.45^{+0.14+0.10}_{-0.46-0.11}$&$0.17^{+0.02+0.01}_{-0.12-0.01}$&0.73&$1.16^{+0.18+0.40}_{-0.37-0.37}$&$0.12^{+0.03+0.03}_{-0.07-0.03}$&0.82\\
$n_{1\mathrm{Mpc}, M_{\star}>10^{10}\,\mathrm{M_{\odot}}} \geq 87$&$2.32^{+0.50+0.56}_{-0.40-0.70}$&$0.12^{+0.06+0.04}_{-0.04-0.06}$&1.52&$1.96^{+0.16+0.08}_{-0.34-0.08}$&$0.24^{+0.04+0.01}_{-0.06-0.02}$&1.60&$3.06^{+0.46+0.33}_{-0.44-0.35}$&$0.12^{+0.06+0.02}_{-0.04-0.02}$&1.11\\
$K_0 < \mathrm{70\,keV\,cm^2}$&$1.60^{+0.15+0.22}_{-0.35-0.22}$&$0.30^{+0.14+0.05}_{-0.06-0.09}$&1.38&$1.46^{+0.09+0.07}_{-0.31-0.08}$&$0.31^{+0.13+0.01}_{-0.05-0.01}$&1.77&$1.89^{+0.25+0.25}_{-0.44-0.27}$&$0.15^{+0.09+0.02}_{-0.03-0.03}$&0.77\\
$K_0 \geq \mathrm{70\,keV\,cm^2}$&$2.38^{+0.55+0.90}_{-0.75-1.02}$&$0.08^{+0.15+0.03}_{-0.02-0.05}$&1.01&$2.00^{+0.23+0.08}_{-0.27-0.08}$&$0.25^{+0.08+0.01}_{-0.02-0.02}$&1.75&$2.84^{+0.49+0.39}_{-0.61-0.51}$&$0.10^{+0.13+0.02}_{-0.01-0.03}$&0.93\\
\hline
\end{tabular}
\end{center}
\end{table*}

\section{A closer investigation of the cluster cores}\label{sec:results2}
In the previous section we discussed the number density and stellar mass density profiles of the ensemble cluster, and found that these are well-described by NFW profiles, except for the inner regions ($R \lesssim 0.10 \cdot R_{200}$). The central parts show a significant and substantial excess, both in galaxy numbers and their stellar mass density distribution. Per cluster this excess within the inner regions is on average $\sim$1 galaxy with $10^{9}<\mathrm{M_\star / M_{\odot} < 10^{10}}$, and $\sim$2 galaxies with $\mathrm{M_\star > 10^{10}\, M_{\odot}}$, and a total stellar mass excess in satellite galaxies of $\sim10^{11}\, \mathrm{M_{\odot}}$ per cluster, compared to the NFW profiles. 

The purple points in Figs.~\ref{fig:number_profile_ensemble} \& \ref{fig:profile_ensemble} show the numbers of spectroscopically confirmed member galaxies. Although these data points are off-set with respect to the full photometric measurement as a result of spec-$z$ incompleteness, they are consistent with the central excess of galaxy numbers and stellar mass density compared to the standard NFW profile. 

To study the central parts of our cluster-sized haloes further, we revisit the fits to the number density and stellar mass density distributions by allowing the inner slope of the density profiles to vary. We hence fit so-called generalised NFW (gNFW) profiles \citep[e.g.][]{zhao96,wyithe01} to the data points. These profiles are described by
\begin{equation}
\rho(r)=\frac{\rho_{0}}{     \left( \frac{r}{R_{s}}  \right)^{\alpha}  \left(1+\frac{r}{R_{s}}\right)^{3-\alpha}   },
\end{equation}
where the concentration is defined, as in the case of the standard NFW profile, to be $c_{\mathrm{NFW}}=\frac{R_{200}}{R{s}}$. For $\alpha =1$, the inner slope equals -1, corresponding to the standard NFW profile. We project the generalised NFW profile numerically along the line-of-sight.

For the number density profiles we find that, for galaxies with $10^{9} < \mathrm{M_\star / M_{\odot}} < 10^{10}$, a profile with $\alpha=1.20^{+0.18+0.03}_{-0.22-0.04}$ and $c=1.41^{+0.42+0.18}_{-0.38-0.12}$ gives a good description of the data ($\chi^2/d.o.f.=1.04$). For the more massive galaxies ($\mathrm{M_\star > 10^{10}\, M_{\odot}}$), the best-fitting parameters are $\alpha=1.64^{+0.04+0.08}_{-0.16-0.06}$ and $c=0.72^{+0.31+0.31}_{-0.19-0.28}$, with goodness-of-fit $\chi^2/d.o.f.=1.19$. Again, both a sample-to-sample variance (first) and systematic (due to cosmic variance in the background, second) error are quoted. The significantly steeper inner slope we find for the high mass sample compared to the lower mass sample indicates that the more massive galaxies are more strongly concentrated in the cluster ensemble. The effect of dynamical friction, which is more efficient for massive galaxies, can be the cause of this mass segregation.

For the stellar mass density we also find a better fit over-all, with $\chi^2/d.o.f.=1.24$ instead of 2.51. The best-fitting profile is given by $\alpha=1.63^{+0.05+0.09}_{-0.25-0.17}$ and $c=0.64^{+0.49+0.73}_{-0.21-0.33}$. The shape of the stellar mass density profile closely agrees with the number density profile for the massive galaxies, which is expected since these dominate in total stellar mass over the less massive galaxies. In Figs.~\ref{fig:number_profile_ensemble} \& \ref{fig:profile_ensemble}, the gNFW profiles are shown by the solid lines. 

For reference we also consider Einasto \citep{einasto65} profiles, which are described by 
\begin{equation}
\rho(r)= \rho_{0}\,\exp \left( -\frac{2}{\alpha} \left[ \left(\frac{r}{R_{s}}\right)^{\alpha} -1 \right]\right),   
\end{equation}
and have been found to provide good fits to the dark matter density distribution of massive haloes in N-body simulations \citep[e.g.][]{dutton14,klypin14}. We project these profiles numerically along the line-of-sight, and find that they give a similarly good representation of the data as the gNFW profile. Parameters $\alpha$ and $c$ ($\equiv\frac{R_{200}}{R{s}}$, as before), and reduced $\chi^2$ values are presented in Table~\ref{tab:einastopars}.

A significant part of the total stellar mass distribution is the stellar mass contained in the BCG and ICL. Although a full accounting of the ICL component is beyond the scope of the current paper, we assess their contribution by measuring the distribution of the stellar mass including the BCG. To measure this total, we directly sum all measured flux around the BCG locations of the original $r$-band images (i.e. without first removing the BCGs' main profiles with {\tt GALFIT}). To estimate the stellar mass distribution, we multiply this with the stellar-mass-to-light ratio (M/L) of the BCG under the assumption that there is no M/L-gradient. We mask the locations of bright stars, sum the flux in annuli that are logarithmically spaced, and statistically subtract the field by considering a large annulus far away from the cluster centres. The background-subtracted central stellar mass density profile is shown as a thick dotted line in Fig.~\ref{fig:profile_ensemble}, with thinner dotted lines marking the 68\% uncertainty region as estimated from cluster-bootstrapping. At a projected radius of $R \sim 0.02 \cdot R_{200}$, the contribution of stellar mass in satellites is roughly similar to that of the BCG component. As a good consistency check we note that the dotted line, which by definition also includes stellar mass in satellites, and the black data points have consistent values in the outermost region where they overlap (at $R \sim 0.08 \cdot R_{200}$). By construction, part of the ICL is also included in this total profile. However, because of the way the background is subtracted from our images, the larger scale component of the ICL is not taken into account. A more sophisticated data reduction is required to measure this component down to sufficiently low surface brightnesses \citep[e.g.][]{NGVS12}, and we leave this to a future study.

\begin{table}
\caption{The excess of stellar mass in satellites in each of the subsamples, with respect to the overall best-fitting NFW profile. Both the relative contribution, and the absolute excess are given, both with respect to the stellar mass included in that NFW profile in the radial regime $R < 0.10 \cdot R_{200}$.}
\label{tab:excess}
\begin{center}
\begin{tabular}{l | c c c }
\hline
\hline
& \multicolumn{3}{c}{Stell mass density}\\
 & & \multicolumn{2}{c}{Central excess}\\
Cluster & $c_{\mathrm{NFW}}$ & Relative & log($\Delta M_\star / M_{\odot}$)\\
\hline
All&$2.03^{+0.20+0.60}_{-0.20-0.40}$&$0.25^{+0.06}_{-0.07}$&$10.95^{+0.09}_{-0.15}$\\
$z<0.114$&$1.56^{+0.46+0.26}_{-0.14-0.18}$&$0.55^{+0.14}_{-0.15}$&$11.20^{+0.10}_{-0.14}$\\
$z\geq0.114$&$2.14^{+0.50+0.78}_{-0.20-0.47}$&$0.18^{+0.08}_{-0.09}$&$10.83^{+0.16}_{-0.28}$\\
$M_{200} < 8.6 \cdot 10^{14}\,\mathrm{M_{\odot}}$&$1.59^{+0.14+0.39}_{-0.36-0.26}$&$0.39^{+0.12}_{-0.12}$&$10.93^{+0.12}_{-0.16}$\\
$M_{200}\geq 8.6 \cdot 10^{14}\,\mathrm{M_{\odot}}$&$2.38^{+0.29+0.67}_{-0.41-0.46}$&$0.15^{+0.08}_{-0.07}$&$10.87^{+0.19}_{-0.26}$\\
$M_\mathrm{{\star,BCG}} < 9.1 \cdot 10^{11}\,\mathrm{M_{\odot}}$&$2.04^{+0.43+0.72}_{-0.37-0.44}$&$0.33^{+0.10}_{-0.11}$&$11.02^{+0.11}_{-0.18}$\\
$M_\mathrm{{\star,BCG}}\geq 9.1 \cdot 10^{11}\,\mathrm{M_{\odot}}$&$1.80^{+0.21+0.38}_{-0.19-0.28}$&$0.28^{+0.09}_{-0.08}$&$11.00^{+0.12}_{-0.16}$\\
$n_{1\mathrm{Mpc}, M_{\star}>10^{10}\,\mathrm{M_{\odot}}} < 87$&$1.43^{+0.06+0.44}_{-0.24-0.27}$&$0.29^{+0.12}_{-0.14}$&$10.75^{+0.15}_{-0.31}$\\
$n_{1\mathrm{Mpc}, M_{\star}>10^{10}\,\mathrm{M_{\odot}}} \geq 87$&$2.38^{+0.44+0.59}_{-0.26-0.38}$&$0.23^{+0.07}_{-0.07}$&$11.07^{+0.12}_{-0.16}$\\
$K_0 < \mathrm{70\,keV\,cm^2}$&$1.65^{+0.20+0.34}_{-0.40-0.23}$&$0.35^{+0.18}_{-0.17}$&$11.05^{+0.18}_{-0.28}$\\
$K_0 \geq \mathrm{70\,keV\,cm^2}$&$2.45^{+0.58+0.64}_{-0.42-0.42}$&$0.27^{+0.09}_{-0.08}$&$11.09^{+0.12}_{-0.16}$\\\hline
\end{tabular}
\end{center}
\end{table}

\subsection{Dependence on cluster physical properties}\label{sec:samplesplits}
Given our sample of 60 clusters over a range of redshifts and halo masses, we investigate if the excess of stellar mass in satellite galaxies is related to any specific cluster (or BCG) property. The properties we consider are cluster redshift and cluster halo mass (see Table~\ref{tab:overview}), BCG stellar mass (based on estimated M/L and integrated $r$-band luminosity with {\tt GALFIT}), and cluster richness\footnote{Defined here as the number of background-subtracted cluster galaxies with $M_{\star}>10^{10}\,\mathrm{M_{\odot}}$ within a projected radius of 1 Mpc from the BCG.}. If we split the sample on the medians of these properties, we measure for each subset a significant central excess in the stellar mass distribution with respect to the best-fitting NFW profile, see Table~\ref{tab:excess}. This excess is $\sim10^{11}\, \mathrm{M_{\odot}}$ per cluster, comprising about 30\% of the stellar mass contained in the NFW profile for $R < 0.10 \cdot R_{200}$. 

A thermodynamical property that is measured for 37 of the clusters in our sample is the central entropy \citep[$K_{0}$, presented in][]{bildfellthesis,mahdavi13}, which is defined as the deprojected entropy profile evaluated at a radius of 20 kpc from the cluster centre. This observable is related to the dynamical state of the cluster \citep{pratt10}, and correlates (by definition) with the inner slope of the gas density distribution. We therefore investigate if the inner part of the stellar mass distribution also depends on this property. \citet{mahdavi13} found a hint of bimodality in the distribution of the central entropy, on either side of $K_0 = \mathrm{70\,keV\,cm^2}$. Following that work, we split our sample between galaxies with central entropies smaller (13 clusters) and higher (24 clusters) than this value. Again, the stellar mass excess is significant in both subsamples (Table~\ref{tab:excess}). 

In each subsample, the gNFW profile provides a better fit to the data than a standard NFW profile. Note that the gNFW profile parameters $\alpha$ and $c$ are degenerate, but none of the splits results in a best-fitting profile that is significantly ($>2\sigma$) different from the over-all stack (Table~\ref{tab:nfwpars}). Note that the splits themselves are not independent of each other, due to relations between, for example, richness and halo mass \citep[e.g.][]{andreon10b}, and a slight covariance between mass and redshift in this sample (Fig.~\ref{fig:evolution_thesis}). Although the stellar mass excess with respect to the NFW profile is thus significant in each subsample, we cannot draw firm conclusions regarding the dependence of the stellar mass profile shape on cluster properties with the current data set.

\section{Discussion - The evolving stellar mass distribution}\label{sec:discussion2}
In this study we found that the NFW profile provides a good description of the stellar mass density distribution of satellites in clusters in the local ($0.04<z<0.26$) Universe, but only at radii R $\gtrsim 0.10\,R_{200}$. Following studies on the evolution of the dark matter profiles in N-body simulations \citep[e.g.][]{duffy08,dutton14}, we discussed the evolution of the stellar mass distribution by considering an evolution in the NFW concentration parameter in Sect.~\ref{sec:discussion1}. However, since Sect.~\ref{sec:results2} shows that there is a significant excess in the stellar mass density distribution of satellite galaxies compared to this NFW profile, a simple comparison of these parameters does not cover the full story. 

Furthermore, note that the concentration parameters we are comparing are defined with respect to the clusters' $R_{200}$. Since the critical density $\rho_{\mathrm{crit}}$, with respect to which these radii are defined, evolves, the measured concentrations will change, even if the physical profile remains constant over time \citep[pseudo-evolution, e.g.][]{diemer13}. Together with the expected halo mass growth of a factor of $\sim 3$ between $z=1$ and $z=0$ \citep{wechsler02,springel05}, $R_{200}$ correspond to a physical size of $\sim 1$ Mpc for the GCLASS clusters, and $R_{200}\sim $ 2 Mpc for the low-$z$ sample.

\begin{figure*}
\resizebox{\hsize}{!}{\includegraphics{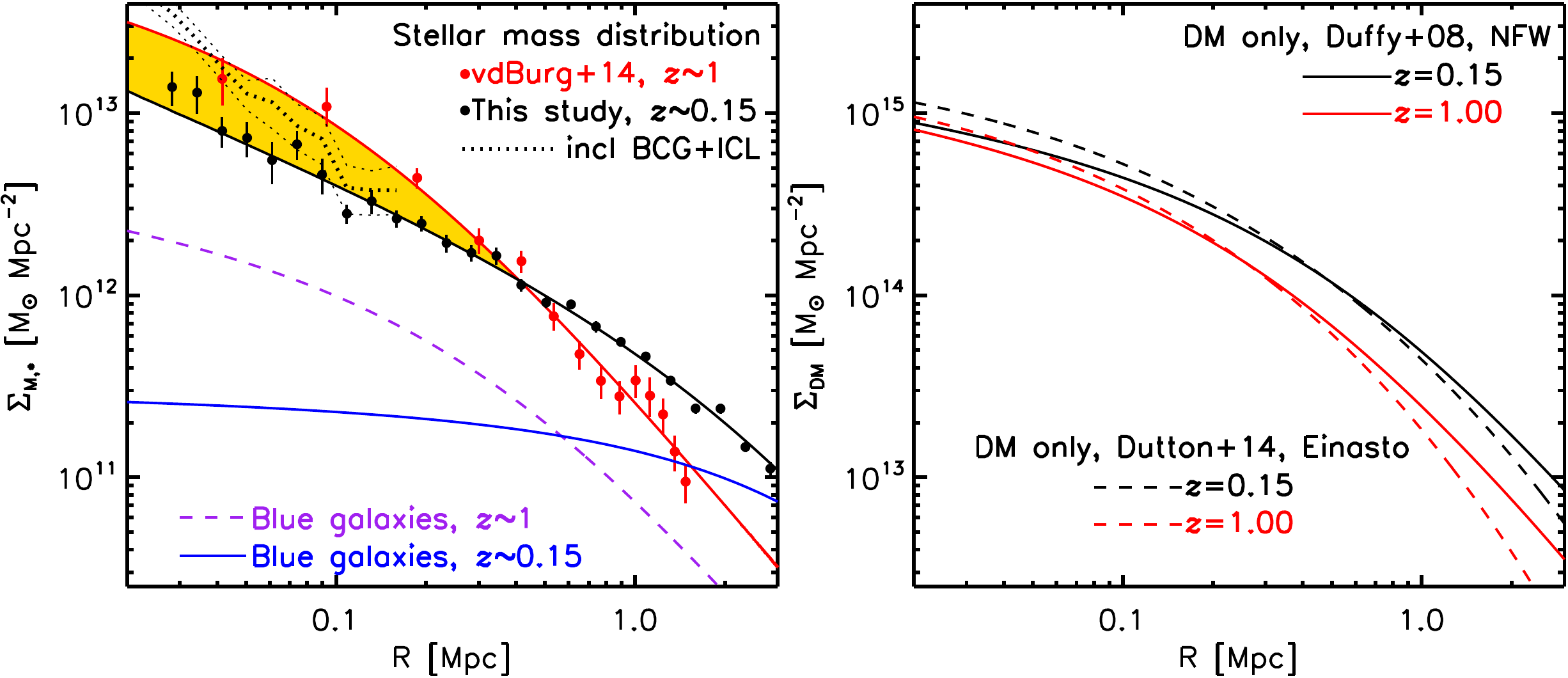}}
\caption{\textbf{Left panel:} \textit{Red points}: The average stellar mass density profile of GCLASS, in physical units. \textit{Black points}: The stellar mass density profile at low-$z$, at the same physical scale. The orange region marks the part of the $z\sim 1$ profile that is in excess of the $z\sim 0.15$ profile, and comprises a stellar mass of about $7\times 10^{11}\,\mathrm{M_{\odot}}$. \textit{Black dotted:} Stellar mass at $z\sim0.15$, including BCG+ICL. \textit{Dashed purple and solid blue:} Stellar mass in blue galaxies at $z\sim1$ and $z\sim 0.15$, respectively. \textbf{Right panel:} Dark matter profiles from N-body simulations, using the average profile parameters from \citet{duffy08} and \citet{dutton14}, but with the profiles plotted on the same physical scale. Shown are profiles at redshifts of $z=1.00$ (\textit{red}) and $z=0.15$ (\textit{black}), with masses of $M_{200}=3\times10^{14}\,\mathrm{M_{\odot}}$ and $M_{200}=9\times10^{14}\,\mathrm{M_{\odot}}$, respectively.} 
\label{fig:insideout_Mpc}
\end{figure*}

Because of these complications, we make a direct, and more intuitive, comparison in Fig.~\ref{fig:insideout_Mpc} (left panel) by studying the cluster stellar mass density profiles \textit{on the same physical scale}. This way we are not affected by pseudo-evolution, are not dependent on a chosen parameterisation of the density profiles, and study directly how the profiles of these clusters evolve since $z\sim 1$. We can make this comparison without re-scaling because the GCLASS clusters are the approximate progenitors of the clusters studied in this work. However, to make a fair comparison we only integrate the stellar mass contained in satellite galaxies down to stellar masses of $\mathrm{M_\star > 10^{10}\, M_{\odot}}$ at $z\sim 0.15$, since that represents the approximate stellar mass depth of GCLASS. 

This purely observational comparison suggests that, although the total stellar mass content of these clusters grows substantially since $z\sim 1$, the stellar mass density in the cluster cores (R $\lesssim$ 0.4 Mpc) is already present at $z=1$. Moreover, there seems to be an excess of stellar mass in this regime at $z\sim 1$ compared to $z\sim 0.15$. Note, however, that in this comparison of the stellar mass in satellite galaxies, we do not take account of the ICL component, and excluded the BCGs from the story. The build-up of stellar mass in these components may explain the observed evolution. Massive galaxies close to the BCG are expected to merge with the central galaxy on a relatively short time-scale, and play a dominant role in the build-up of stellar mass in the BCG \citep[e.g.][]{burke13,lidman13}. The stellar mass contained between the two curves in Fig.~\ref{fig:insideout_Mpc} (left panel, orange region), is on average $\sim 7\times 10^{11}\,\mathrm{M_{\odot}}$ per cluster. Given that the BCGs in the GCLASS clusters have typical stellar masses of $M_{\star,\mathrm{BCG}} \simeq 3\times 10^{11}\,M_{\odot}$ (vdB14, Table~2), and that the median stellar mass contained in the BCGs in the sample studied here is $\mathrm{M_\star \simeq 9\times 10^{11}\, M_{\odot}}$, it is an interesting coincidence that this excess of stellar mass in satellites at $z\sim1$ roughly equals the difference in BCG stellar mass between the two samples. The development of an ICL component may also contribute to an evolution in the observed stellar mass density profile. The dotted line in Fig.~\ref{fig:insideout_Mpc} (left panel) was already shown in Fig.~\ref{fig:profile_ensemble}, and is consistent with the picture that stellar mass reassembles itself in the direction of the central galaxy, becoming part of the BCGs' extended light profiles.

A sufficient amount of stellar mass that is required for BCG growth thus seems to already be present in the centres of the clusters at $z\sim 1$, although it is still part of the satellite galaxy population. However, while these satellites seem to drive most of the BCG mass growth, it is interesting why they do not get replenished with new infalling satellites. In the more massive haloes at low-$z$, the process of dynamical friction, which is supposed to effectively reduce the orbital energy of massive infalling satellites, seems to work less efficiently. This might be related to the observational result that the massive end of the stellar mass function hardly evolves over cosmic time \citep[e.g.][]{muzzin13b}, whereas the haloes we study grow in mass by a factor $\sim 3$. Compared to $z>1$, the time that it takes for a massive galaxy to lose enough orbital energy to arrive at the centre is longer in the local universe.

On the other hand, substantial growth of the stellar mass content in the cluster outskirts ($R \gtrsim 1.0\, \mathrm{Mpc}$) is required to match the low-$z$ descendants of the GCLASS systems. Under the assumption that galaxies populate sub-haloes and that these systems are accreted onto the clusters since $z=1$, it is expected that dark-matter haloes also accrete matter onto the outskirts. This effect is indeed observed in N-body simulations, if these simulations are compared on the same physical scale, see Fig.~\ref{fig:insideout_Mpc} (right panel). Recently, \citet{dutton14} and particularly \citet{klypin14} have shown that Einasto profiles provide a better description of the dark matter density distribution of massive haloes in N-body simulations. As a comparison, we therefore compare the results of \citet{duffy08}, which are based on a WMAP5 cosmology and an NFW parameterisation, with those from \citet{dutton14}, which are based on a \textit{Planck} cosmology and an Einasto parameterisation. Both are normalised to have the same $M_{200}$. In both cases, the profiles of these simulated haloes grow at all radii, although their growth is smaller in the centre. The evolution between the observed stellar mass distribution and the dark matter in N-body simulations is thus significantly different (cf. Fig.~\ref{fig:c_evolution_new}), independent of the used parameterisation. 

The observations strongly suggest a scenario in which the stellar mass component grows in an inside-out fashion, indicating that the presence of baryons plays an important role in this assembly process. The observed evolution of the stellar mass distribution is thus a stringent test for existing and future hydrodynamical simulations \citep[e.g.][]{schaye10,cen14,vogelsberger14,genel14,schaye15}, as it is of importance both in a cosmological context and in our quest to understand the formation and evolution of galaxies in our Universe.

\subsection{Radial distribution of different galaxy types}
\begin{figure}
\resizebox{\hsize}{!}{\includegraphics{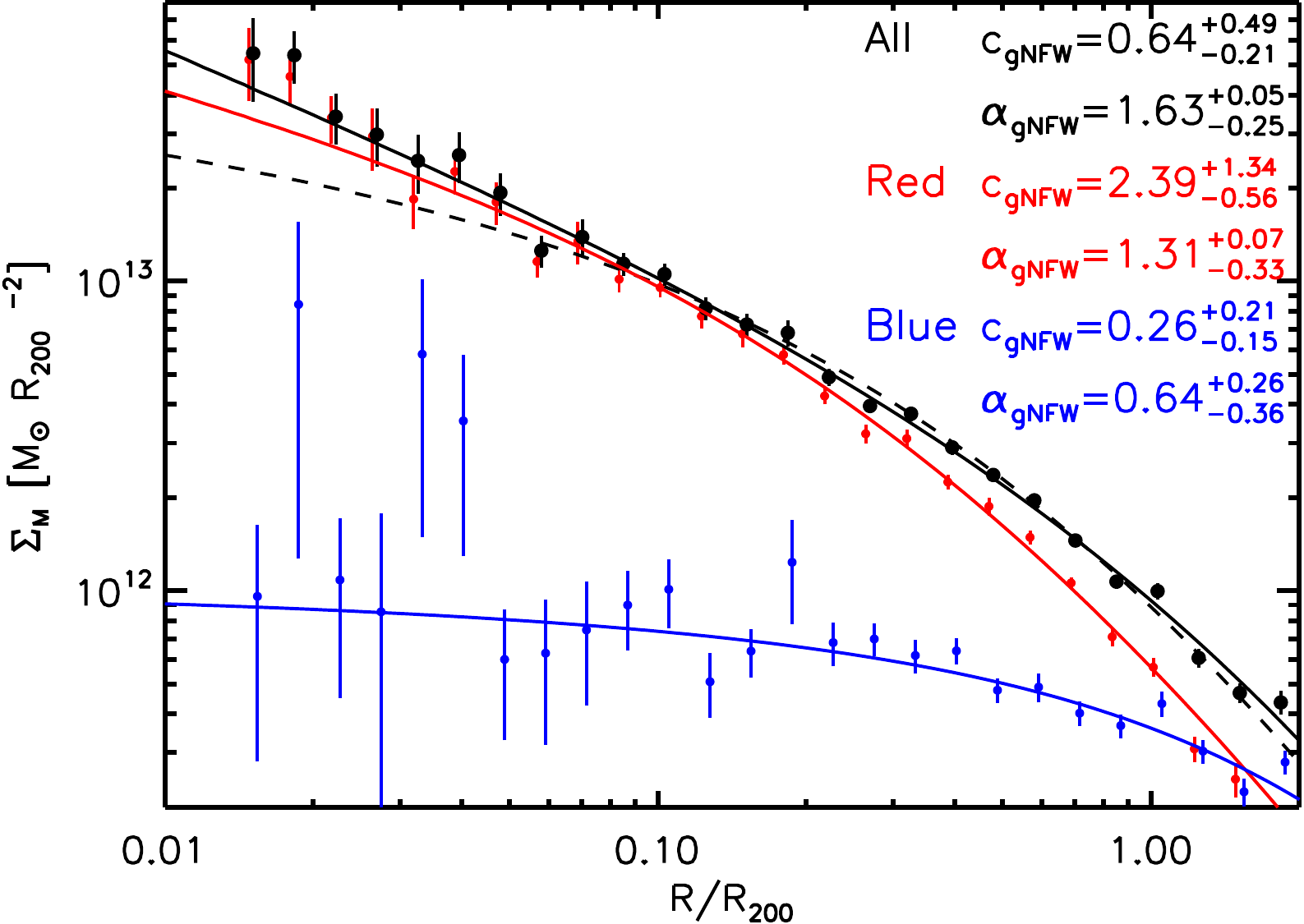}}
\caption{\textit{Black:} The stellar mass density profile of Fig.~\ref{fig:profile_ensemble}, separated between red-sequence \textit{(red)} and bluer \textit{(blue)} galaxies. The best-fitting gNFW profiles to the red and blue sub-samples are also shown here.}
\label{fig:galtypes}
\end{figure}
Since blue galaxies are thought of as a dynamically younger component of the galaxy cluster population than red galaxies, a distinction between galaxy types can yield further insight in the way clusters accrete their satellite population. We make a distinction between red and blue galaxies using as simple criterion the cluster red sequence in the ($g-r$)-colour. Since the colour of the red sequence is redshift-dependent, we identify the red-sequence population in each of the individual clusters, and stack the resulting stellar mass distributions in Fig.~\ref{fig:galtypes}. The best-fitting gNFW profile to each of the galaxy types is plotted.

Figure~\ref{fig:galtypes} shows that galaxies on the red sequence completely dominate the stellar mass distribution in the cluster centres, and are dominant over bluer galaxies (in terms of stellar mass density) up to at least $R_{200}$. Bluer galaxies are still significantly over-dense compared to the field over the entire radial range that is shown (the field values are subtracted), but note the shallow inner slope of the gNFW profile that describes the blue galaxies ($\alpha_{\mathrm{gNFW}}=0.64^{+0.26}_{-0.36}$). 

In Fig.~\ref{fig:insideout_Mpc} (left panel), we show the blue galaxy population of this sample, and also the blue galaxy population in the GCLASS clusters (vdB14, their Fig.~7). This shows that there is a dramatic evolution in the relative radial distribution of blue galaxies compared to the overall galaxy population. The blue fraction of cluster galaxies is lower overall at low-$z$ compared to high-$z$, but the difference is most prominently visible near the cluster cores. In the highly-simplified picture that blue galaxies fall in, and quench by some environmental process with a delay of several Gyr \citep[e.g.][]{wetzel2013,muzzin14}, we can use their locations in the cluster to study where the stellar mass is most recently accreted. Even after more than a dynamical time-scale, which is typically 1 Gyr, the blue galaxies are mostly on the outskirts of the clusters (note that we are studying the \textit{projected} surface mass density here). Although the physics involved in the quenching of galaxies require a more detailed modelling, this simplified picture supports a scenario in which clusters assemble their stellar mass distribution in an inside-out fashion. We leave a more detailed discussion on the relative distributions of blue and red-sequence galaxies to a future paper, in which we measure and discuss the stellar mass functions for each of these populations.

\section{Summary and conclusions}\label{sec:conclusion}
In this paper we perform a detailed study of the radial galaxy number density and stellar mass density distribution of satellites in a sample of 60 massive clusters in the local Universe ($0.04<z<0.26$). The cluster sample we study is close in halo mass to the likely descendant population of the $z\sim 1$ GCLASS cluster sample (vdB14), for which a stellar mass concentration of $c_{\mathrm{NFW}}=7.12^{+1.53}_{-0.99}$ was estimated. The main conclusions of this study at low-$z$, and the comparison with GCLASS, can be summarised as follows.

\begin{itemize}
\item The number density and stellar mass density distribution of satellites in this sample are well-described by an NFW profile in the radial range R $\gtrsim 0.10\, R_{200}$. The estimated NFW concentration parameters are roughly consistent with literature measurements with which we can compare.
\item At smaller radii, there is a significant and substantial excess in the amount of stellar mass in satellites of about $\sim 10^{11}\,\mathrm{M_{\odot}}$ per cluster, compared to the standard NFW profile. The central parts (R $\lesssim 0.10\, R_{200}$) of both the number density and stellar mass density distributions are thus significantly steeper than NFW, but generalised NFW profiles with steep inner slopes, and Einasto profiles fit these distributions well. We do not find a significant correlation between the central excess and cluster properties such as redshift, cluster mass, BCG stellar mass, galaxy richness, and cluster central entropy. 
\item A naive comparison between the NFW concentration parameter at $z=1$ from GCLASS, and the measurement at $z\sim 0.15$ suggest a dramatic evolution over cosmic time. This observed evolution in the stellar mass concentration is significantly different from what is found for the dark matter distribution in N-body simulations. 
\item For a more intuitive and direct comparison, we study the stellar mass density distributions between the two epochs on the same physical scale, showing that the stellar mass density in satellites in the cluster cores (R$<$0.4 Mpc) is already in place at $z= 1$; There is a sufficient amount of stellar mass present for the BCGs to grow by a factor of 3. 
\item Although the centre of the stellar mass distribution seems to be in place by $z=1$, substantial growth onto the cluster outskirts is required for the two samples to connect. This inside-out growth picture is quantitatively different from the behaviour of dark matter in N-body simulations, illustrating the importance of baryonic effects in understanding the accretion of smaller haloes by galaxy clusters.
\end{itemize}

\begin{acknowledgements}
We thank Rob Crain, Gabriel Pratt, Monique Arnaud, and Herv\'e Aussel for valuable discussions during the course of this project. This work is based on observations made with the Isaac Newton Telescope through program IDs I10AN006, I10AP005, I10BN003, I10BP005, I11AN009, I11AP013. We thank Malin Velander, Emma Grocutt, Lars Koens and Catherine Heymans for help in acquiring the data. The Isaac Newton Telescope is operated on the island of La Palma by the Isaac Newton Group in the Spanish Observatorio del Roque de los Muchachos of the Instituto de Astrof\'isica de Canarias. Based on observations obtained with MegaPrime/MegaCam, a joint project of CFHT and CEA/DAPNIA, at the Canada-France-Hawaii Telescope (CFHT) which is operated by the National Research Council (NRC) of Canada, the Institute National des Sciences de l'Univers of the Centre National de la Recherche Scientifique of France, and the University of Hawaii.

The research leading to these results has received funding from the European Research Council under the European Union's Seventh Framework Programme (FP7/2007-2013) / ERC grant agreement n$^{\circ}$ 340519. RvdB and HH acknowledge support from the Netherlands Organisation for Scientific Research grant number 639.042.814. CS acknowledges support from the European Research Council under FP7 grant number 279396. MLB acknowledges support from an NSERC Discovery Grant, and NOVA and NWO grants that supported his sabbatical leave at the Sterrewacht at Leiden University. 
\end{acknowledgements}

\bibliographystyle{aa} 
\bibliography{MasterRefs} 

\begin{appendix}

\section{Selection effects in GCLASS}\label{sec:gclassselection}
Given the significant evolution that is observed between the GCLASS sample and the low-$z$ descendant sample, we have to consider the possibility that this inferred evolution is caused by the way these samples are selected. Since it is impossible to select a cluster sample based on halo mass, different selection methods (X-ray, SZ-, or galaxy selections) potentially result in a biased sample of clusters. 

The GCLASS sample consists of ten clusters drawn from the 42 degree Spitzer Adaptation of the Red-sequence Cluster Survey \citep[SpARCS,][]{muzzin09,wilson09,demarco10}. Clusters in SpARCS were detected using the red-sequence detection method developed by \citet{gladdersyee00}, and expanded on in \citet{muzzin08}. In summary, this detection method was applied to the optical+InfraRed data in SpARCS, so that the $z'-3.6\mu\mathrm{m}$ colour was used to detect clusters at redshifts $z>0.8$ after convolving the galaxy number density maps with an exponential kernel \citep[see][Eq.~3]{gladdersyee00}. Richnesses were measured in fixed apertures with a radius of 500 kpc, after which the richest systems were considered for follow-up photometry and spectroscopy. \citet{muzzin12} describes how this GCLASS follow-up sample was drawn from the richest systems after optimising the redshift baseline and ensuring a spread in RA for observational convenience. The fixed aperture of 500 kpc makes the richness selection independent on concentration. However, in principle it is possible that richness and concentration are correlated quantities, such that a richness selection indirectly biases our sample towards high/low concentrations. 

\begin{figure}
\resizebox{\hsize}{!}{\includegraphics{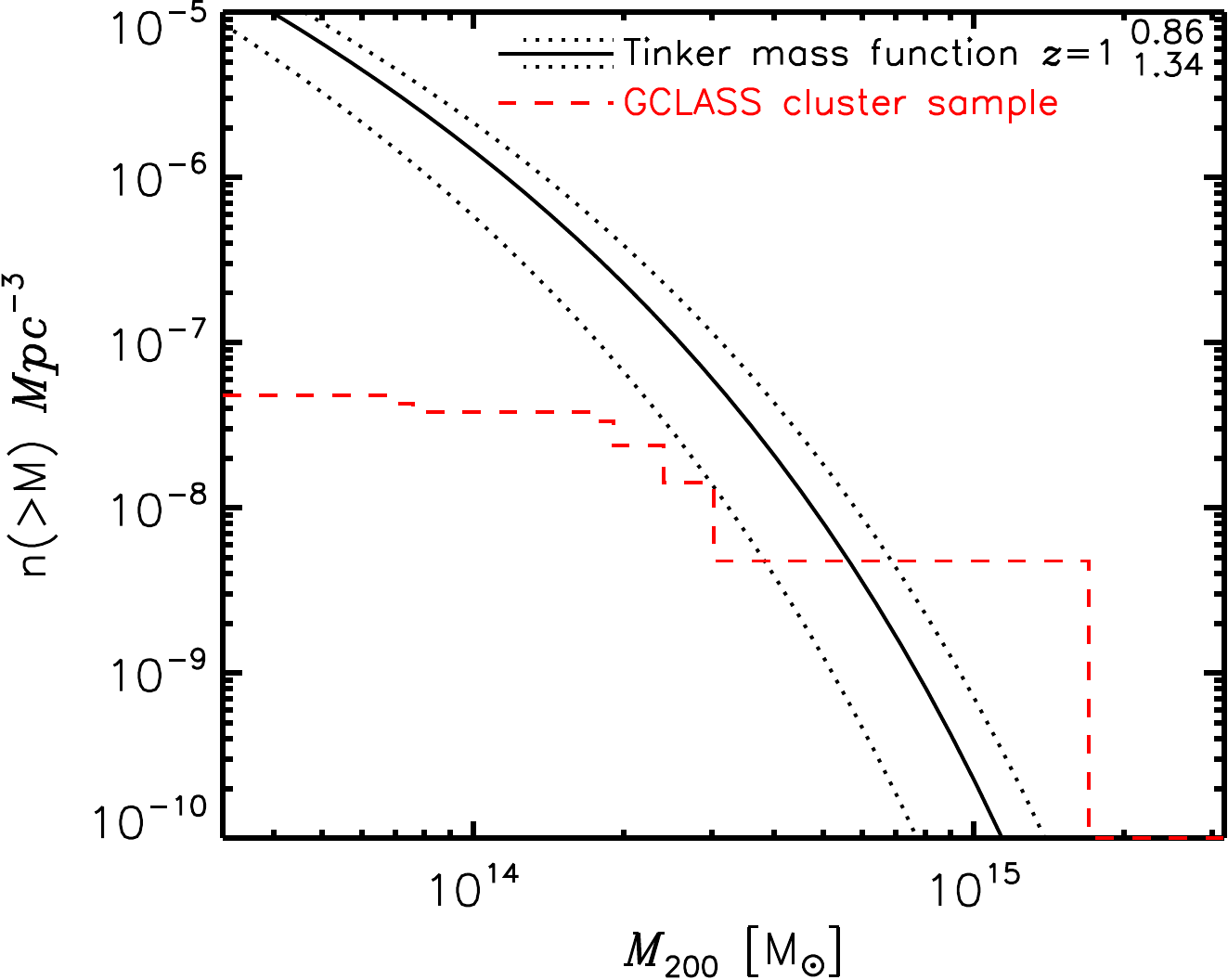}}
\caption{\textit{Black solid line}: \citet{tinker08} cumulative mass function at $z=1$ for WMAP7 cosmology. \textit{Black dotted lines}: \citet{tinker08} cumulative mass functions at $z=0.86$ and $z=1.34$, which are the redshift limits within which the GCLASS clusters are selected. \textit{Red dashed line}: cumulative mass function of the ten GCLASS clusters, normalised by the total volume of SpARCS.}
\label{fig:massfunction}
\end{figure}

The statistics in the GCLASS sample are insufficient to study a possible trend between richness and concentration at $z\sim 1$, but we proceed to test a potential bias in the selection of GCLASS by comparing the dynamical masses of the GCLASS sample to the \citet{tinker08} cumulative halo mass function based on a WMAP7 cosmology, which we show in Fig.~\ref{fig:massfunction}. 
\begin{figure}
\resizebox{\hsize}{!}{\includegraphics{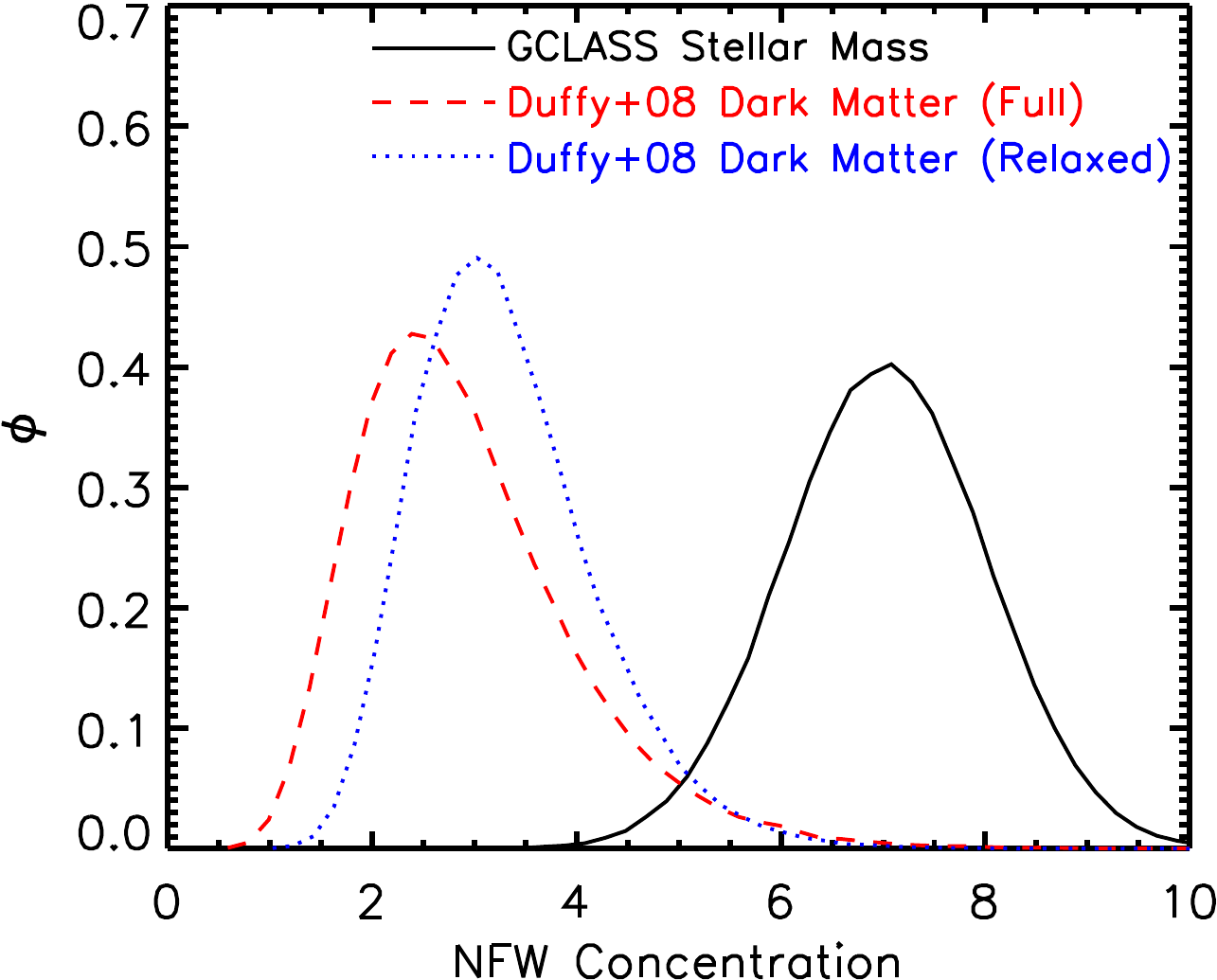}}
\caption{\textit{Black solid line}: GCLASS ensemble average stellar mass concentration with a Gaussian probability distribution around $c=7.12$. Also shown are the log-normal concentration distribution for clusters with the same mass and redshift as the GCLASS sample for the relaxed haloes in \citet{duffy08} (\textit{Blue dotted line}), and their full sample (\textit{Red dashed line}).}
\label{fig:conccomp}
\end{figure}
Given the effective area of 41.9 square degrees we estimate the effective volume of the SpARCS survey (from which GCLASS was selected) in the redshift slice $0.86<z<1.34$ and normalise the cumulative number density of the GCLASS clusters over this volume. At the high-mass end of the distribution we expect Poisson scatter, and there is scatter in the mass-richness relation to be considered. The ten GCLASS systems are therefore not necessarily the most massive ones. Based on this comparison, we estimate that in GCLASS we probe around 10\% of the clusters in the SpARCS volume around the median mass of the GCLASS sample ($M_{200}\simeq 10^{14.3}\,\mathrm{M_{\odot}}$). 

We consider the possibility that the clusters probed by GCLASS are the 10\% with the highest concentrations in the simulation. Figure~\ref{fig:conccomp} shows the GCLASS ensemble average stellar mass concentration with a Gaussian probability distribution around $c=7.12$. The \citet{duffy08} log-normal concentration distribution for cluster-sized haloes in N-body simulations are also shown, both for their relaxed and full sample (haloes were categorised based on the distance between the most bound particle and the centre of mass in the simulation). The relaxed sample has a slightly higher concentration of $c$=3.30 compared to $c$=2.84 for the full sample, but has a smaller scatter ($\sigma (\mathrm{log_{10}}c)$=0.11 dex versus 0.15 dex for the full sample). Where the \citet{duffy08} distributions overlap with the GCLASS probability distribution, these two distributions are similar. 

We perform a simple test in which we randomly sample 100 concentrations from the \citet{duffy08} relations. We do this for 1000 different realisations and each time average the ten most concentrated ones. In only 3\% of the realisations we do find a larger average than the measured concentration from GCLASS ($c=7.12^{+1.53}_{-0.99}$), taking account also of the error on this measured concentration. Therefore, even under the most conservative assumption that a richness selection is completely biased towards the most concentrated galaxy clusters, there is only a 3\% probability that we measure an average concentration for GCLASS of $c=7.12^{+1.53}_{-0.99}$. Moreover, as we argued in vdB14, the measured concentration of $c\simeq 7.12$ is a lower limit if we include uncertainties arising from offsets between the BCGs and the "true" cluster centres. Given these arguments, it is unlikely that both the observed evolution since $z\sim 1$, and the difference between the predictions from N-body simulations and observations at this redshift, are only an effect of the way the GCLASS sample is selected. 

\end{appendix}
\end{document}